\documentclass[hyper]{prop2015}% no class options needed by now

\usepackage{bbm}

\def\bec{\begin{center}}
\def\ec{\end{center}}

 \def\det{{\rm det\,}}
\def\be{\begin{equation}}
\def\ee{\end{equation}}
\def\bea{\begin{eqnarray}}
\def\eea{\end{eqnarray}}
\def\ba{\begin{array}}
\def\ea{\end{array}}

\usepackage{color}
\definecolor{red}{rgb}{1.5,0,0}
\definecolor{lred}{rgb}{0.3,0,0}
\definecolor{green}{rgb}{0,1,0}
\definecolor{blue}{rgb}{0,0,1}
\definecolor{violet}{rgb}{0.8,0,0.8}

\category{Proceedings}
\title{Higher Gauge Structures  in Double and Exceptional Field Theory}
\subtitle{\href{http://www.maths.dur.ac.uk/lms/109/index.html}{LMS/EPSRC Durham Symposium on Higher Structures in M-Theory}}
\author[O.~Hohm]{Olaf Hohm\inst{a,}\footnote{Corresponding author e-mail:~\href{mailto:ohohm@physik.hu-berlin.de}{\textsf{ohohm@physik.hu-berlin.de}}}}
\author[H.~Samtleben]{Henning Samtleben\inst{b}}
\address[1]{Institute for Physics, Humboldt University Berlin,\\
 Zum Gro\ss en Windkanal 6, D-12489 Berlin, Germany}
\address[2]{Univ Lyon, Ens de Lyon, Univ Claude Bernard, CNRS,\\
Laboratoire de Physique, F-69342 Lyon, France}
\shortauthors{O.~Hohm, H.~Samtleben}
\begin{abstract}
We review the higher gauge symmetries  in double and exceptional field theory from 
the viewpoint  of an embedding tensor construction. This is based on a (typically infinite-dimensional) 
Lie algebra $\frak{g}$ and a choice of representation $R$. 
The embedding tensor is a map from the representation space $R$ into $\frak{g}$ 
satisfying a compatibility condition (`quadratic constraint'). The Lie algebra structure on 
$\frak{g}$ is transported to a Leibniz--Loday algebra on $R$, which in turn gives rise to an 
$L_{\infty}$-structure. We review how the gauge structures of double and exceptional field theory fit into this 
framework. 
\end{abstract}
\shortabstract
\begin{document}
\maketitle
%%% Use this if the article text won't start with a \section:
% \noindent
%%% Being based on LaTeX's article class, and2012 supports the respective 
%%% sectioning level from \section to \subparagraph.

\section{Introduction}

Our goal in this article is to review double and exceptional field theory 
\cite{Siegel:1993th,Hull:2009mi,Hull:2009zb,Hohm:2010jy,Hohm:2010pp,Hillmann:2009ci,Berman:2010is,Coimbra:2011ky,Hohm:2013pua,Hohm:2013vpa,Hohm:2013uia,Hohm:2014fxa}, 
which are T- and U-duality covariant 
formulations of (low-energy limits or truncations of) string  and M-theory, with a particular 
emphasis on their higher gauge structures going beyond Lie algebras. These are particularly  encoded 
in so-called tensor hierarchies: towers of $p$-form gauge fields  transforming under 
non-Abelian gauge symmetries. The higher gauge algebra of double field theory was originally 
derived from closed string field theory \cite{Hull:2009mi}, which itself is governed by a higher gauge algebra, 
a Lie-infinity (or $L_{\infty}$) algebra \cite{Zwiebach:1992ie}. In contrast, the gauge structures of exceptional field theory, and most notably 
their tensor hierarchies, were first  constructed on a case-by-case basis that obscures some 
of the unifying features. Recently such more unifying approaches have emerged, in which the higher algebras of exceptional field theory
are treated more systematically and, in particular, 
are derived from a Lie algebra and an `embedding tensor' map \cite{Hohm:2018ybo}. 
This requires the generalization of techniques developed in gauged supergravity to infinite-dimensional Lie algebras based on function spaces. 
We use the opportunity to present and streamline this new viewpoint in a self-contained fashion. 
Related and complementary accounts include \cite{Baraglia:2011dg,Deser:2014mxa,Deser:2016qkw,Cederwall:2018aab,Cagnacci:2018buk,Arvanitakis:2018cyo}.

\subsection{General approach}
We start by presenting Einstein's theory of general relativity in a 
somewhat unfamiliar fashion that, however, sets the stage for our subsequent generalizations to (the low-energy effective actions of) 
string and M-theory. Pure Einstein gravity in $D$ dimensions is defined by the action 
 \be\label{DEinstein}
  S \ = \ \int {\rm d}^DX\sqrt{{ G}}\,R({ G})\;, 
 \ee
with the metric tensor ${G}_{\hat{\mu}\hat{\nu}}$, where $\hat{\mu},\hat{\nu}=1,\ldots, D$. 
The idea is now to perform a $D=n+d$ split, assuming that the $D$-dimensional space permits a suitable foliation, 
but without any further topological assumptions and without truncating physical degrees of freedom. 
We then decompose indices and coordinates, writing for the coordinates 
$X^{\hat{\mu}}=(x^{\mu},y^m)$ (to which we refer to as external and internal coordinates, respectively)
and for the metric components 
 \be\label{Gdecomp}
  \begin{split}
   G_{\mu\nu} \ &= \ |G|^{-\frac{1}{n-2}} g_{\mu\nu} + G_{mn} \,A_{\mu}{}^{m} A_{\nu}{}^{n}\;, \\
   G_{\mu m} \ &= \ G_{mn}\, A_{\mu}{}^{n}\;,
  \end{split}
 \ee
where $G_{mn}$ is the internal $d\times d$ block of $G$, and $|G|\equiv \det G$. 
We emphasize that here the fields are still assumed to depend on all $n+d$ coordinates. 
The ansatz (\ref{Gdecomp}) thus does not entail any truncation; we have merely parameterized 
the metric components in a convenient fashion. 
Inserting (\ref{Gdecomp}) into (\ref{DEinstein}) one obtains 
 \be\label{splitEH}
  \begin{aligned}
   S \ &= \ \int {\rm d}x\, {\rm d}y\sqrt{g}\,\Big(\widehat{R}(g) - \tfrac{1}{4}|G|^{\frac{1}{n-2}} G_{mn}\, F^{\mu\nu m} F_{\mu\nu}{}^{n}\,+\\
   &\kern2.5cm+\tfrac{1}{4} D^{\mu} G^{mn} D_{\mu}G_{mn}\,-\\
   &\kern2.5cm- \tfrac{1}{4(n-2)}\big(|G|^{-1}D_{\mu}|G|\big)^2 -V(G,g)\Big)\;, 
  \end{aligned}
 \ee
where $V(G,g)$ is a function involving only `internal' derivatives $\partial_m$ (including the Ricci scalar of $G_{mn}$).
In order to explain the various terms in (\ref{splitEH}), let us recall that this action is a rewriting of the Einstein--Hilbert action 
(\ref{DEinstein}) and hence must be invariant under $D$-dimensional diffeomorphisms, including the internal transformations 
 \be\label{intdiff}
  y^m \ \rightarrow \ y^{\prime m} \ = \ y^m - \xi^m(x,y)\;. 
 \ee
The `vector' fields $A_{\mu}{}^m(x,y)$ transform under this symmetry as a connection, 
 \be
  \delta_{\xi}A_{\mu}{}^{m} \ = \ D_{\mu}\xi^m \ \equiv \ \partial_{\mu}\xi^m -{\cal L}_{A_{\mu}}\xi^{m}\;, 
 \ee
where we defined covariant derivatives and introduced the notation of Lie derivatives that generate infinitesimal internal
diffeomorphisms: on a generic vector $W^m$ we have 
 \be\label{LieFirst}
  {\cal L}_{V}W^m \ \equiv \ \big[V,W\big]^m \ = \ V^n\partial_n W^m - W^n\partial_n V^m\;, 
 \ee
and similar formulas hold for arbitrary tensor fields. The action (\ref{splitEH})
is written in terms of these covariant derivatives $D_{\mu}$. In particular, the `covariantized' Ricci scalar $\widehat{R}$
is obtained by replacing $\partial_{\mu}\rightarrow D_{\mu}$ in the familiar definitions. 
Moreover, the vector fields are governed by the non-Abelian field strengths 
 \be
  F_{\mu\nu}{}^{m} \ = \ \partial_{\mu}A_{\nu}{}^{m} -\partial_{\nu}A_{\mu}{}^{m} -\big[A_{\mu}, A_{\nu}\big]^m\;,  
 \ee
with the Lie bracket (\ref{LieFirst}). The complete action (\ref{splitEH}) is thus manifestly invariant under internal diffeomorphisms (\ref{intdiff}).

Naturally, the formulation (\ref{splitEH}) is the ideal starting point for Kaluza--Klein compactifications 
to $n$ dimensions. For a torus reduction, for instance,  one declares the fields to be independent of the $d$ internal 
coordinates to obtain 
  \be\label{KKEH}
  \begin{split}
   S \ = \ \int {\rm d}x  &\sqrt{g}\,\Big({R}(g) - \tfrac{1}{4}|G|^{\frac{1}{n-2}} G_{mn}\, F^{\mu\nu m} F_{\mu\nu}{}^{n}+\\
   &\kern.8cm+\tfrac{1}{4} \partial^{\mu} G^{mn} \partial_{\mu}G_{mn} - \tfrac{1}{4(n-2)}\big(|G|^{-1}\partial_{\mu}|G|\big)^2\Big)\;, 
  \end{split}
 \ee
where now all derivatives are partial derivatives and $F_{\mu\nu}{}^{m}=2\,\partial_{[\mu}A_{\nu]}{}^{m}$ is 
the Abelian $U(1)^d$ field strength. 
The idea is now to reinterpret (or reconstruct) the full theory (\ref{splitEH})  as a non-Abelian generalization 
or `gauging' of an intermediate Abelian theory. This intermediate theory is 
obtained from (\ref{KKEH}) by promoting all fields to depend arbitrarily on the $y$ coordinates but \textit{without} introducing 
derivatives $\partial_m$ in the action. The action then takes the same form as (\ref{KKEH}), 
   \be\label{intermediateEH}
  \begin{split}
   S \ = \ \int &{\rm d}x {\rm d}y \sqrt{g}\,\Big({R}(g) - \tfrac{1}{4}|G|^{\frac{1}{n-2}} G_{mn}\, F^{\mu\nu m} F_{\mu\nu}{}^{n}+\\
   &+\tfrac{1}{4} \partial^{\mu} G^{mn} \partial_{\mu}G_{mn} - \tfrac{1}{4(n-2)}\big(|G|^{-1}\partial_{\mu}|G|\big)^2\Big)\;, 
  \end{split}
 \ee  
but with the difference that all fields depend on $(x,y)$ and that there is an additional $y$-integration. 
Taking the internal dimensions to be compact, say a torus, this theory can be thought of 
as a decompactification limit of (\ref{splitEH}) in the  following sense. Upon expanding all fields into $y$-space Fourier modes,  
each internal derivative is multiplied by $\frac{1}{R}$, where $R$ is a characteristic length scale of 
the internal space (as the radius of a circle). Then sending $R\rightarrow \infty$ decouples all internal derivatives 
$\partial_m$, leaving a theory for `massless' fields with action (\ref{intermediateEH}).\footnote{For the special case $n=1$, the external dimension 
being time, this phenomenon is central to the BKL analysis of spacelike singularities, where close to the singularity 
all spatial gradients (internal derivatives) decouple, c.f.~the discussion in \cite{Nicolai:2005su}.}

The above discussion shows that (\ref{splitEH}) is a consistent deformation of the `unbroken phase' (\ref{intermediateEH}), 
where the deformation is governed by the finite parameter $\frac{1}{R}$. 
In this review we will emphasize a point of view that starts from the `global' 
(i.e.\ $x$-independent)
symmetries 
of this unbroken phase and promotes a certain subalgebra  to a gauge symmetry. 
In order to explain  this approach we first have to examine  the symmetries of (\ref{intermediateEH}). 
The local symmetries are given by $n$-dimensional diffeomorphisms with parameters $\xi^{\mu}(x,y)$ and $U(1)^d$ 
gauge symmetries with parameters $\xi^m(x,y)$, where the parameters can depend arbitrarily on $y$ since there are no $\partial_m$ derivatives 
that could detect this dependence. A perhaps  unconventional feature of this intermediate theory  (\ref{intermediateEH}) is that in addition it 
has two types of independent global symmetries. First, we have a ${\rm GL}(d)$ invariance acting on indices $m,n,\ldots$
Indeed, we can think of (\ref{intermediateEH}) as a non-linear sigma model based on ${\rm GL}(d)/{\rm SO}(d)$, with the additional feature that the parameters 
can be $y$-dependent, hence giving rise to an infinite-dimensional extension of the symmetry. Second, we have a global internal diffeomorphism symmetry 
of the $y^m$ coordinates. Summarizing, the global symmetries are 
 \be\label{globalPARAMeters}
  \begin{split}
   {\rm GL}(d)\;\; {\rm transformations}\,&: \quad \sigma_{m}{}^{n}(y)\;, \\
   \text{internal diffeomorphisms}\,&: \quad \lambda^m(y)\;. 
  \end{split}
 \ee
For instance, on the internal metric $G_{mn}$ these symmetries act as\footnote{It seems more natural to postulate a transformation 
w.r.t.~$\lambda^m$ given by the full Lie derivative, but in presence 
of the independent $GL(d)$ symmetry with parameters $\sigma_m{}^n$ this is equivalent to the above form modulo  parameter redefinitions.} 
 \be\label{gREP}
  \delta G_{mn} \ = \ \lambda^k\partial_k G_{mn} +2\,\sigma_{(m}{}^{k} G_{n)k}\;,
 \ee
and on the vector fields $A^m$ as 
 \be\label{deltaAINTRO}
  \delta_{\zeta}A^m \ = \ \lambda^n\partial_nA^m - A^n\sigma_{n}{}^m \;. 
 \ee
Denoting the parameters (\ref{globalPARAMeters}) collectively as\linebreak $\zeta=(\lambda^m, \sigma_m{}^{n})$, these 
transformations close according to the bracket 
 \be\label{globalLIE}
  \big[\zeta_1,\zeta_2\big] =  2\big(\lambda_{[1}{}^{n}\partial_n\lambda_{2]}{}^m, \,\lambda_{[1}{}^{k}\partial_k \sigma_{2]m}{}^n
  +\sigma_{[1 m}{}^{k} \sigma_{2]k}{}^{n}\big)\,.
 \ee
This bracket satisfies the Jacobi identity and hence defines an (infinite-dimensional) Lie algebra $\frak{g}$. 
A closely related Lie algebra should be familiar to most physicists: the semi-direct sum of the algebra of one-dimensional 
diffeomorphisms (the Witt or Virasoro algebra) with the affine extension of any 
Lie algebra. 

Our goal is now to gauge a certain subalgebra of (\ref{globalLIE}), by which we mean that we want to promote the resulting parameters 
to become $x$ dependent (in addition to the $y$ dependence that we here think of as parametrizing  the `global' symmetry 
algebra). The one consistent gauging that we know, corresponding to our 
starting point of $D$-dimensional Einstein gravity, suggests that only a subalgebra can be gauged, in which $\lambda^m$ is identified with internal 
(but $x$-dependent) diffeomorphisms and $\sigma$ with derivatives of the same diffeomorphism parameter. 
More formally, we can describe this gauging of (\ref{intermediateEH})
in terms of a map from the representation space $R$ of vector fields 
into the Lie algebra $\frak{g}$. This map $ \vartheta\,: \;R \,\rightarrow\, \frak{g}$, which is referred to as the embedding tensor,  
 is in the present case given by 
 \be\label{varthetaintro}
  \vartheta(\xi) \ \equiv \ \big(\xi^m, \,\partial_m\xi^n\big) \ \in \frak{g}\;. 
 \ee
 
To understand the significance of this map, let us note the general fact that the $R$ representation of $\frak{g}$ 
is the representation in which the vector fields $A_{\mu}{}^{m}$ transform. This is so because the vector fields of 
the ungauged (unbroken) phase are to be used for the gauging. In the case at hand, this 
$\frak{g}$  representation acts on the vector as (\ref{deltaAINTRO}). 
With this and (\ref{varthetaintro}) one infers that the original Lie derivatives 
(\ref{LieFirst}) describing infinitesimal (internal) diffeomorphisms of full general relativity are recovered as 
 \be
  {\cal L}_{\xi}A^m \ \equiv \ \delta_{\vartheta(\xi)}A^m\;. 
 \ee
Similarly, the covariant derivatives are recovered by the gauging $\partial_{\mu}\rightarrow D_{\mu}\equiv \partial_{\mu}-\delta_{\vartheta(A_{\mu})}$, 
as are the non-Abelian field strengths. The consistency of the procedure moreover requires that external diffeomorphism transformations under  
$x^{\mu}\rightarrow x^{\mu}-\xi^{\mu}(x,y)$ get suitably deformed, which in turn necessitates the introduction of the `scalar potential' 
$V(g,G)$ in (\ref{splitEH}). Since we know that the final answer (\ref{splitEH}) is a rewriting of Einstein gravity and hence consistent 
we do not have to elaborate further on this, although it could be illuminating to investigate the (presumably remote) possibility 
that there are other consistent gaugings that do not lead back to $D$-dimensional Einstein gravity.

The above reconstruction of general relativity  from an `unbroken' phase and an embedding tensor map may seem 
like an overly formal presentation of a well-known theory, but it turns out that this general viewpoint illuminates several 
features of double and exceptional field theory that otherwise appear rather ad-hoc. In the remainder of this section 
we briefly illustrate some of these features. 
We begin with the low-energy effective action of bosonic string theory (or the NS-NS sector of 
superstring theory), 
featuring the metric $G$, Kalb--Ramond 2-form $B$ and scalar dilaton $\phi$, 
 \be\label{NSNSaction}
  S \ = \ \int {\rm d}^DX\, e^{-2\phi}\big(R(G)+4(\partial\phi)^2-\tfrac{1}{12}H^2\big)\;, 
 \ee
where $H={\rm d}B$. Decomposing the coordinates according to a $D=n+d$ split, 
and truncating the dependence on the internal coordinates $y^m$, one naturally obtains an 
$n$-dimensional theory with a global ${\rm GL}(d)$ invariance as in (\ref{KKEH}). 
However, this theory actually exhibits a larger (hidden) symmetry  given by the non-compact group $O(d,d)$, 
with the action given by \cite{Maharana:1992my}
 \be\label{firstActionIntro0}
  \begin{split}
   S \ = \ & \int {\rm d}x\, \sqrt{g}\,e^{-2\phi}\Big(  {R}
   +4\partial^{\mu}\phi \partial_{\mu}\phi -\tfrac{1}{12}{\cal H}^{\mu\nu\rho}{\cal H}_{\mu\nu\rho}+\\
   &+\tfrac{1}{8}\partial^{\mu}{\cal H}^{MN}\partial_{\mu}{\cal H}_{MN}
   -\tfrac{1}{4}{\cal H}_{MN}{F}^{\mu\nu M}{F}_{\mu\nu}{}^{N}\Big)\;. 
  \end{split}
 \ee  
Here the fields are organized into $O(d,d)$ representations (whose fundamental indices are $M, N=1,\ldots, 2d$), namely: 
singlets $g_{\mu\nu}$, $b_{\mu\nu}$ and $\phi$, a vector field $A_{\mu}{}^M$ with field strength $F_{\mu\nu}{}^M=2\,\partial_{[\mu}A_{\nu]}{}^{M}$, 
and the symmetric tensor 
 \be\label{firstH0}
  {\cal H}_{MN} \ = \  \begin{pmatrix}    G^{ij} & -G^{ik}B_{kj}\\[0.5ex]
  B_{ik}G^{kj} & G_{ij}-B_{ik}G^{kl}B_{lj}\end{pmatrix} \ \in \ O(d,d)\;, 
 \ee
defined in terms of the internal metric and $B$-field. Thus, (\ref{firstActionIntro0}) is manifestly $O(d,d)$ invariant. 

We can extend this theory to an `unbroken phase' of the original theory 
so that its fields depend on $d$ internal coordinates, with a global symmetry algebra (\ref{globalPARAMeters}) that merges 
$d$-dimensional internal diffeomorphisms with ${\rm GL}(d)$ transformations. 
By construction, this theory can be deformed so as to reconstruct (\ref{NSNSaction}). 
However, since (\ref{firstH0}) is $O(d,d)$ invariant it is more natural to define the `unbroken phase' by introducing 
$2d$ coordinates  $Y^M$ so that the global symmetry algebra is defined as in (\ref{globalPARAMeters}) 
but with $2d$-dimensional diffeomorphisms merging with $O(d,d)$ transformations. 
Moreover, since there are actually $2d$ vector fields $A_{\mu}{}^{M}$ available for the gauging 
one may then hope that there is a deformation that  preserves $O(d,d)$. 
Indeed, the needed doubling of coordinates is 
precisely what is required by string theory on toroidal backgrounds, where (\ref{NSNSaction}) is incomplete, featuring 
massive Kaluza-Klein (momentum) modes (associated to the $y^m$ dependence) but missing dual so-called winding modes. 
The winding modes are naturally encoded (for instance in closed string field theory \cite{Kugo:1992md}) in the dependence on dual coordinates $\tilde{y}_m$, 
merging with the original coordinates into an $O(d,d)$ vector $Y^M=(\tilde{y}_m, y^m)$. 
The resulting `double field theory' is then invariant under the discrete subgroup $O(d,d;\mathbbm{Z})$ that preserves the periodicity 
conditions of the torus.

In contrast to the (re-)construction of general relativity, an important new feature arises for double field theory: 
while the resulting gauge algebra of general relativity is a Lie algebra, the 
gauge algebra of double field theory turns out to be a higher algebra 
with higher brackets. This is a generic feature that persists for exceptional field theory, where the global 
(U-duality) symmetry group belongs to the series E$_{d(d)}$, $2\leq d \leq 9$, (which has the T-duality group $O(d-1,d-1)$ as a 
subgroup), and where the enhanced theory features coordinates in the fundamental representation of E$_{d(d)}$.
The higher algebraic structures are partly due to the presence of constraints on the coordinate dependence (the so-called section constraints), 
which in string theory is a manifestation of the level-matching constraints. More generally, the emergence of higher 
algebraic structures can be understood as a consequence of the fact that one attempts to transport an algebraic structure,  
the Lie algebra $\frak{g}$, to a different space, the representation space $R$ in which the vector fields live. 
Since generically these spaces are \textit{not} isomorphic, the Lie algebra structure 
is not transported to a Lie algebra structure, but rather to a Lie-infinity ($L_{\infty}$) algebra. 
In the mathematics literature it is well established  that under `homotopy equivalences' algebraic structures 
can be transported to `infinity' versions of the same structure (see, e.g., the `derived bracket' construction in \cite{tv:higherder,Getzler:1010.5859}), but 
the embedding tensor formulation seems not to be widely known, and we hope that the present review may remedy this.

The rest of this article is organized as follows. In Section~\ref{sec:2} we introduce the embedding tensor formalism in an invariant (index-free) 
fashion that makes it applicable to infinite-dimensional Lie algebras, and we briefly discuss how the resulting higher algebras 
give rise to $L_{\infty}$-algebras. 
In Section~\ref{sec:3} we show how the generalized diffeomorphisms of double and exceptional field theory 
can be obtained from an embedding tensor construction based on an infinite-dimensional Lie algebra (that is the global 
symmetry algebra of an `unbroken phase' as outlined above). The general construction is then applied to the special 
cases of duality groups $O(d,d)$, E$_{7(7)}$ and E$_{8(8)}$.  
Finally, in Section~\ref{sec:4}, we turn to the construction of `tensor hierarchies', in which the higher gauge structures manifest themselves 
in the presence of higher $p$-form gauge fields, and we use these to give complete dynamical equations encoding 
in particular 11-dimensional or type IIB supergravity. In the conclusion Section~\ref{sec:5} we discuss open problems.

\section{Higher algebras via embedding tensors}\label{sec:2}
\label{sec:higher}

\subsection{Embedding tensor}\label{embeddingtensorsec}

We start with a (finite- or infinite-dimensional) Lie algebra $\frak{g}$ with Lie brackets $[\cdot,\cdot]$, whose 
elements  are denoted by small Latin or Greek letters $a,b, \ldots$ or $\zeta, \xi,\ldots$, respectively. 
Any Lie algebra is equipped with a representation, the adjoint representation on $\frak{g}$, defined 
by the familiar infinitesimal transformation 
\be
 \delta_{\zeta}a \ \equiv \ {\rm ad}_{\zeta}a \ \equiv \ [\zeta, a]\;. 
\ee
Moreover, there is a \textit{coadjoint} representation on the dual space $\frak{g}^*$, whose elements we denote 
by calligraphic letters ${\cal A}, {\cal B}$. The coadjoint action, denoted by 
 \be
   \delta_{\zeta}{\cal A} \ \equiv \ {\rm ad}^*_{\zeta}{\cal A}\;, 
 \ee
is defined so that the pairing of vectors and covectors, 
 \be
  {\cal A}(a) \ \in \ \mathbbm{R}\;, 
 \ee
is invariant: 
 \be\label{dualinvariance}
  \delta_{\zeta}({\cal A}(a)) \ \equiv \ ({\rm ad}^*_{\zeta}{\cal A})(a) + {\cal A}({\rm ad}_{\zeta}(a)) \ = \ 0\;. 
 \ee

Consider now an arbitrary $\frak{g}$ representation on a vector space $R$, whose elements we denote 
by capital Latin or Greek letters from the middle of the alphabet. Being a  representation, we have infinitesimal transformations 
on vectors $V\in R$, denoted by 
 \be\label{Roperators}
  \delta_{\zeta}V \ = \ \rho_{\zeta}V\;, 
 \ee
where the operators $\rho_{\zeta}$ satisfy 
 \be\label{representation}
  [\rho_{\zeta_1},\rho_{\zeta_2}]  \ = \ \rho_{[\zeta_1,\zeta_2]} \;. 
 \ee
It should be emphasized that on the left-hand side $[\cdot,\cdot]$ denotes the commutator of operators, and on the right-hand side 
it denotes the original (abstract) Lie bracket. 
This representation also has a dual representation on the dual space $R^*$, whose elements we denote by capital letters $A, B, \ldots$
from the beginning of the alphabet. Being the dual space, there is a pairing between vectors and covectors, $A(V) \in  \mathbbm{R}$, that 
is invariant under the combined $\frak{g}$ action: 
denoting the  operators acting on $R^*$ by $\rho_{\zeta}^*$ we have, in parallel to (\ref{dualinvariance}), 
 \be\label{copropertyyyy}
  (\rho_{\zeta}^* A)(V) \ = \ -A(\rho_{\zeta}V)\;. 
 \ee 

Our goal is now to transport the Lie algebra structure on $\frak{g}$ to an algebraic structure on $R$ by means 
of an embedding tensor map 
 \be\label{embeddingtensormap}
  \vartheta\,: \; R\;\rightarrow\; \frak{g}\;. 
 \ee
For the special case that $R$ is equivalent to the adjoint representation, we can take $\vartheta$ to be an isomorphism (for instance, the identity 
map if $R = \frak{g}$), in which case $R$ trivially inherits the Lie algebra structure of $\frak{g}$. 
This underlies the standard construction of non-Abelian Yang-Mills theory.
In general, however, the space $R$ may be larger or smaller 
than $\frak{g}$, so that $\vartheta$ cannot be an isomorphism. Thus, the Lie algebra structure on $\frak{g}$ generally cannot be transported to a Lie algebra structure on $R$. 
We will now impose a constraint (`quadratic constraint' \cite{deWit:2002vt}), that implies that $R$ inherits a \textit{higher} algebra structure. This higher algebra is a Leibniz--Loday algebra, which in turn yields a so-called \textit{strongly homotopy} Lie algebra 
or $L_{\infty}$-algebra \cite{Kotov:2018vcz}.  

In order to state the quadratic constraint we note that the embedding tensor map (\ref{embeddingtensormap}) and the representation (\ref{Roperators}) 
yield a natural bilinear algebraic structure on $R$, defined for $V,W\in R$ by 
 \be\label{GENLeibnizZZ}
  V \circ W \ \equiv \ \rho_{\vartheta(V)}W\;. 
 \ee 
This `product' $\circ$, which in general is not antisymmetric, 
defines an action of $R$ on itself by $\delta_VW\equiv V\circ W$. The most direct way to state the quadratic constraint 
is to demand that the commutator of this action closes. This means that for $U,V,W\in R$
 \be\label{LeibnizZERO}
  V\circ (W\circ U)-W\circ (V\circ U) \ = \  (V\circ W)\circ U\;. 
 \ee 
Algebras with a bilinear operation satisfying this relation are known as Leibniz (or Loday) algebras.  
The alternative writing of this relation given by 
 \be\label{Leibnizproperty}
   V\circ (W\circ U) \ = \ (V\circ W)\circ U +  W\circ (V\circ U) \;
 \ee
makes clear where the name Leibniz algebra comes from: the `adjoint' action defined by 
$\circ$ acts according to the Leibniz rule on the same product.   
In the case that this operation is antisymmetric, the above relations reduce to the Jacobi identity for Lie algebras, 
and hence Lie algebras are special cases of Leibniz--Loday algebras.

The relations (\ref{LeibnizZERO}) (or (\ref{Leibnizproperty})) represent the quadratic constraints that the 
embedding tensor (\ref{embeddingtensormap}) has to satisfy, but below it will be beneficial to 
provide alternative forms of this  constraint from which (\ref{LeibnizZERO}) can be derived. 
To this end we re-interpret the embedding tensor as a map
 \be
  \Theta\,:\; R\,\otimes \, \frak{g}^*\; \rightarrow \; \mathbbm{R}\;, 
 \ee 
where $\frak{g}^*$ is the dual space to $\frak{g}$. 
For $V\in R$ and  ${\cal A}\in \frak{g}^*$ this map is defined by 
 \be\label{THETAbig}
  \Theta(V,{\cal A}) \ \equiv \ -{\cal A}(\vartheta(V)) \;,  
 \ee
using the pairing between vector and covector on the right-hand side. (We introduced a sign for later convenience.)  
The claim is that invariance of $\Theta$, i.e., 
 \be\label{ThetaINV}
  \delta_{\Lambda}\Theta(V,{\cal A}) \ \equiv \ \Theta(\Lambda\circ V,{\cal A})+\Theta(V,{\rm ad}_{\vartheta(\Lambda)}^*{\cal A}) \ = \ 0\;, 
 \ee
implies the Leibniz algebra relations. Since in examples it is typically easier to verify that the `scalar' $\Theta$ is invariant, 
(as opposed to verifying  `vector' relations such as  (\ref{Leibnizproperty})), this observation will be crucial for our 
applications below. 

In order to prove this claim we first note for any representation $R$ there is a canonical map 
 \be
  \pi\,:\; R \otimes  R^* \; \rightarrow\; \frak{g}^*\;, 
 \ee  
defined as follows: since its image is a coadjoint vector, it naturally acts on adjoint vectors $\zeta$, 
and so we can define, for $V\in R$, $A\in R^*$,
  \be\label{PIDEF}
  (\pi(V,A))(\zeta) \ \equiv \ (\rho^*_{\zeta}A)(V)\;. 
 \ee
This map is convenient because the Leibniz product (\ref{GENLeibnizZZ}) can then be written, upon pairing with a covector, as
 \be\label{INVEMBEDDDDDD}
  A(V\circ W) \ = \ \Theta(V, \pi(W,A))\;. 
 \ee
This relation is proved as follows: 
 \be
 \begin{split}
  A(V\circ W) \ &= \ A(\rho_{\vartheta(V)}W) \ = \  
  -(\rho^*_{\vartheta(V)}A)(W) \\
  \ &= \ -(\pi(W,A))(\vartheta(V))
  \ = \ \Theta(V,\pi(W,A))\;, 
 \end{split} 
 \ee 
where we used (\ref{copropertyyyy}) and (\ref{PIDEF}).  
Let us next note that the map $\pi: R\otimes R^*\rightarrow \frak{g}^*$ was defined using only invariant maps, which implies that it 
transforms `covariantly': 
 \be\label{natrualnessofpi}
  \delta_{\zeta}(\pi(V,A)) \ \equiv \ \pi(\rho_{\zeta}V,A)+\pi(V,{\rho}_{\zeta}^*A) \ = \ {\rm ad}_{\zeta}^*(\pi(V,A))\;. 
 \ee 
Invariance of $\Theta$ then implies invariance of the left-hand side of (\ref{INVEMBEDDDDDD}): 
  \be\label{invariancePairinggg}
  \begin{split}
   \delta_{\Lambda}(A(V\circ W)) \ &= \ \delta_{\Lambda}\Theta(V,\pi(W,A)) 
    \ = \ 0\;. 
  \end{split} 
  \ee 
In here we can now write out the left-hand side as follows  
 \be
 \begin{split}
  0 \ &= \ (\rho_{\vartheta(\Lambda)}^*A)(V\circ W) + 
  A((\Lambda\circ V)\circ W+V\circ (\Lambda\circ W)) \\
  \ &= \ A(-\Lambda\circ (V\circ W) +(\Lambda\circ V)\circ W+V\circ (\Lambda\circ W))\;. 
 \end{split} 
 \ee
Thus, we obtained the Leibniz algebra relations upon pairing with a (co-)vector $A\in R^*$.   
Since this holds for arbitrary $A$, and we assume the usual non-degeneracy condition 
 \be\label{nondegeneracy}
  \forall {A}\ \in \ {R}^*\,:\;\; {A}(V) \ = \ 0\quad \Rightarrow\quad V \ = \ 0\;, 
 \ee 
 the Leibniz relations follow.

We close this subsection by presenting a convenient alternative form of the 
quadratic constraint. To this end we note that the invariance condition (\ref{ThetaINV}) reads 
by means of (\ref{THETAbig})
\be\label{pairedquadconstr}
\begin{split}
 0 \ &= \ {\cal A}(\vartheta(\Lambda\circ V)) + ({\rm ad}_{\vartheta(\Lambda)}^*{\cal A})(\vartheta(V)) \\
 \ &= \ {\cal A}(\vartheta(\Lambda\circ V)) - {\cal A}({\rm ad}_{\vartheta(\Lambda)}\vartheta(V)) \\
 \ &= \ {\cal A}\big(\vartheta(\Lambda\circ V)-[\vartheta(\Lambda),\vartheta(V)]\big)\;, 
\end{split}
\ee
where we used (\ref{dualinvariance}). Since this holds for arbitrary ${\cal A}$, we
conclude, assuming the analogue of (\ref{nondegeneracy}),  that the expression that is paired with ${\cal A}$ has to vanish. 
We then infer from (\ref{pairedquadconstr}) that 
for any $V,W\in R$
 \be\label{finalquadrconstr2346}
 \vartheta(V\circ W)   \ = \ [\vartheta(V), \vartheta(W)] \;. 
 \ee
It then immediately follows  that
any $\frak{g}$-representation $R$ with infinitesimal action $\delta_\zeta$
lifts to a representation of the Leibniz algebra via $\delta_{\vartheta(\Lambda)}$: 
 \be
  [\delta_{\vartheta(\Lambda_1)}, \delta_{\vartheta(\Lambda_2)}] \ = \ \delta_{[\vartheta(\Lambda_1),\vartheta(\Lambda_2)]} \ =  \ 
  \delta_{\vartheta(\Lambda_1\circ \Lambda_2)}\;.
 \ee

\subsection{$L_{\infty}$-algebras}\label{Linfsection}

We will now relate the `higher' algebras discussed above to strongly homotopy Lie algebras ($L_{\infty}$-algebras) 
\cite{Zwiebach:1992ie,Lada:1992wc,Hohm:2017pnh}. 
Our main goal here is to connect to the known higher algebras in the literature as a way of a brief pedagogical introduction
to  $L_{\infty}$-algebras. The content of this subsection will not play a prominent role in the remainder of this paper, but 
we will use the opportunity to introduce some useful notation.

Our starting point is a general Leibniz algebra with a `product' $\circ$ satisfying (\ref{Leibnizproperty}). 
For our present discussion we do not have to assume that this algebra is derived from an underlying Lie algebra 
by an embedding tensor construction. We recall that any Leibniz algebra has a natural action on itself, given by 
 \be
  \delta_{V}W \ \equiv \ {\cal L}_{V}W \ \equiv \ V\circ W\;, 
 \ee
where we introduced the notation ${\cal L}$, below to be used for generalized Lie derivatives.  
These transformations close in that 
 \be\label{gaugealgebra}
 \begin{split}
  [{\cal L}_V, {\cal L}_W]U \ & \equiv \  {\cal L}_V({\cal L}_W U)- {\cal L}_W({\cal L}_V U) \\
  \ &= \ V\circ (W\circ U) -W\circ (V\circ U) \\
   \ &= \ (V\circ W)\circ U \\
    \ &= \ {\cal L}_{V\circ W}U\;, 
 \end{split}
 \ee 
using the Leibniz relation (\ref{LeibnizZERO}).  
Next, defining 
 \be\label{symmantisymm}
 \begin{split}
  \{V,W\} \ &\equiv \ \tfrac{1}{2}(V\circ W + W\circ V)\;, \\
  [V,W] \  &\equiv  \ \tfrac{1}{2}(V\circ W - W\circ V)\;, 
 \end{split} 
 \ee 
and (anti-)symmetrizing (\ref{gaugealgebra}) in $V, W$ we have 
 \be\label{genCLosure}
  [{\cal L}_V,{\cal L}_W]U  
  \  = \ {\cal L}_{[V,W]}U\;, 
 \ee
and   
 \be\label{symmistrivial}
  {\cal L}_{\{V,W\}}U \ = \ 0 \quad \forall V,W \;. 
 \ee
Thus, the antisymmetric part defines the more conventional algebra, 
but there is in general a notion of `trivial parameters', given by the symmetric part. 

Using only the general relations above we can now prove that the `Jacobiator' of the bracket $[\cdot\,,\cdot]$ is trivial in 
the sense of being writable in terms of $\{\cdot, \cdot\}$, 
 \be\label{Jacobiator}
  {\rm Jac}(V_1, V_2, V_3) \ \equiv \ 3[[V_{[1}, V_2],V_{3]}] \ = \ \{V_{[1}\circ V_{2},  V_{3]}\}\;. 
 \ee
For the proof we suppress the total antisymmetrization in the three arguments. We then need to establish: 
 \be
  6[V_1\circ V_2, V_3] - 2\{V_1\circ V_2, V_3\} \ = \ 0\;, 
 \ee
where we multiplied by 2 for convenience.  
We then write out the brackets and use  total antisymmetry: 
 \be
  \begin{split}
  &6[V_1\circ V_2, V_3] - 2\{V_1\circ V_2, V_3\}  \\
   & \ = \, 3(V_1\circ V_2)\circ V_3 - 3\,V_3\circ (V_1\circ V_2)\,- \\
  & \ \quad  -(V_1\circ V_2)\circ V_3 - V_3\circ (V_1\circ V_2) \\
  & \ = \ 2\,(V_1\circ V_2)\circ V_3-4\,V_3\circ (V_1\circ V_2)\\
  & \ = \ 2\,(V_1\circ V_2)\circ V_3+2\, V_2\circ (V_1\circ V_3) - 2\, V_1\circ (V_2\circ V_3)  \\
  & \ = \ 0\;,   
  \end{split}
 \ee     
applying  the Leibniz identity (\ref{LeibnizZERO}) in the last step. 

In order to connect to $L_{\infty}$-algebras it is convenient to introduce 
a new notation by writing the symmetric bracket in terms of a linear map ${\cal D}$ and a 
new symmetric operation $\bullet$ as 
 \be\label{DREL}
  \{V,W\}  \ \equiv \ \tfrac{1}{2}{\cal D}(V\bullet W)\;.  
 \ee
This form can be assumed without loss of generality, since without further 
specification ${\cal D}$ can be taken to be the identity and $\bullet$ as a new notation. 
However, in non-trivial examples ${\cal D}$ will emerge naturally as an operator onto a subspace of 
the Leibniz algebra. In turn, this operator could have a non-trivial kernel. 
As a consequence of (\ref{symmistrivial}), we can always choose ${\cal D}$
such that its image is entirely contained within the space of trivial parameters.
Using this notation, the Jacobiator (\ref{Jacobiator}) takes the form
 \be\label{JacobiatorEXPr}
  {\rm Jac}(V_1,V_2,V_3) \ = \ \tfrac{1}{2}{\cal D}([V_1,V_2]\bullet V_3)\;. 
 \ee
Moreover, it is a simple exercise to verify 
 \be\label{idealprop}
  [U, \{V,W\} ] \ = \ \{ U\circ V, W\} + \{V,U\circ W\} - \{U,  \{V,W\} \}\;, 
 \ee
which with (\ref{DREL}) implies that the bracket $[U,{\cal D}(V\bullet W)]$ is 
`${\cal D}$ exact'. We assume that this holds for any argument $f$ in the space 
in which $\bullet$ takes values, so that we can write 
 \be\label{idealbracket}
  [V,{\cal D}f] \ = \ {\cal D}(V(f))\;, 
 \ee
where $V(f)$ is defined implicitly by this relation, up to 
contributions in the kernel of ${\cal D}$.

Let us now turn to $L_{\infty}$-algebras. They are defined on a vector 
space 
  \be
   X \ = \  \bigoplus_{n \in \mathbbm{Z} }  X_n 
 \ee  
with integer grading.  Moreover, $X$ is a \textit{chain complex}, which means that 
it is equipped with a nilpotent differential ${\ell_1}$ of intrinsic degree $-1$, 
mapping between the spaces as 
\be\label{exactsequence}
  \cdots \; \rightarrow \; X_1 \; \xrightarrow{\ell_1} \; X_0 \; \xrightarrow{\ell_1} 
  \; X_{-1}\;
  \rightarrow \; \cdots 
 \ee
The $L_{\infty}$ structure is given by a (potentially infinite) series of linear maps or brackets
$\ell_n: X^{\otimes n}\rightarrow X$, $n=1,2,3\ldots$ In our conventions, these brackets 
have intrinsic degree $n-2$, which means that the degree of their output is the sum of the degree of 
all arguments plus $n-2$. (In this discussion we restrict ourselves to  arguments  with 
definite degrees.) Moreover, the $\ell_n$ are graded antisymmetric, which means that the 
exchange of two adjacent arguments gives a sign unless both arguments have odd parity. 

Most importantly, the $\ell_n$ are subject to a (potentially) infinite number of quadratic 
identities, which replace (and are the `homotopy version' of) the Jacobi identities of Lie algebras. 
Somewhat symbolically, they are given by 
 \be
  \sum_{i+j=n+1} (-1)^{i(j-1)}\ell_j\ell_i \ = \ 0\;, 
 \ee
for each $n=1,2,3\ldots$ These relations can be given a precise mathematical meaning by interpreting the 
$\ell_i$ as coderivations on a suitable tensor algebra, but rather than discussing this in more detail 
here we content ourselves with giving the explicit relations for $n=1,2,3$. For $n=1$ the generalized 
Jacobi identity simply reads $\ell_1^2=0$.
For $n=2$ one obtains, for arbitrary arguments $x_1,x_2\in X$, 
  \be
  \label{L2L1}
 \ell_1(\ell_2(x_1,x_2)) \ = \ \ell_2(\ell_1(x_1),x_2) + (-1)^{x_1}\ell_2(x_1,\ell_1(x_2))\;, 
 \ee
which states that $\ell_1$ acts as a derivation on the `2-bracket' $\ell_2$. 
Finally, for $n=3$ the generalized Jacobi identity reads 
 \begin{equation}\label{LinftyJacobi}
\begin{split}
  0    \  =  \ &\;
  \ell_2(\ell_2(x_1,x_2),x_3) +\text{2 terms} \\[0.5ex]
 &\;+ \ell_1(\ell_3 (x_1,x_2, x_3))  \\[0.5ex]
&\;   + \ell_3(\ell_1 (x_1) ,x_2, x_3) 
  +\text{2 terms} \;, 
  \end{split} 
\end{equation}
where the first line is the (graded) Jacobiator. We thus learn that for $L_{\infty}$-algebras 
the naive Jacobi identity can be violated. The failure of the Jacobi identity is then related to 
the failure of $\ell_1$ to act as a derivation on the `3-bracket' $\ell_3$, given by the second and 
third line.

We now return to the bracket induced by a Leibniz algebra and show how it defines 
an $L_{\infty}$-algebra. This follows from a general result in \cite{Hohm:2017cey} and has also been discussed in 
\cite{Kotov:2018vcz}, although the $L_{\infty}$ extension relevant below 
typically needs to be more general. Postponing a general treatment to future work 
here we focus on the first few relations, without worrying whether there 
may be an obstruction at higher level. Concretely,  we restrict ourselves 
to the part of the chain complex given by 
\be\label{SIMPLEexactsequence}
  X_1 \; \xrightarrow{\ell_1} \; X_0\;, 
 \ee
taking $X_0$ to be the vector space of the Leibniz algebra, and $\ell_1={\cal D}$ to be the operator 
defined implicitly by (\ref{DREL}), with $X_1$ the image of $\bullet: X_0\otimes X_0\rightarrow X_1$.  
The $n=1$ relations $\ell_1^2=0$ trivialize on the truncated complex (\ref{SIMPLEexactsequence}). 
The $n=2$ relations (\ref{L2L1}) for arguments $V\in X_0$ and $f\in X_1$ require 
 \be
  {\cal D}(\ell_2(V,f))  \ = \ \ell_2(V,{\cal D}f) \ = \ {\cal D}(V(f))\;, 
 \ee
where we used that ${\ell_1}(V)=0$ on the complex (\ref{SIMPLEexactsequence}) since there is 
no space $X_{-1}$. Moreover, we used (\ref{idealbracket}) in the last step. 
We infer that this relation is satisfied for 
 \be
  \ell_2(V,f) \ = \ -\ell_2(f,V) \ = \ V(f)\;. 
 \ee
Finally, we turn to the $n=3$ relations (\ref{LinftyJacobi}), which for all arguments in $X_0$  
reads
 \be
  {\rm Jac}(V_1,V_2,V_3) + {\cal D}(\ell_3(V_1,V_2,V_3)) \ = \ 0 \;. 
 \ee
Comparing with (\ref{JacobiatorEXPr}) we infer that this relation is satisfied for the 3-bracket 
 \be
  \ell_3(V_1,V_3,V_3) \ = \ -\tfrac{1}{2}[V_{[1}, V_{2}]\bullet V_{3]} \ \in \ X_1\;, 
 \ee
where we reinstated  the explicit total antisymmetrization. Note that this $\ell_3$ takes 
values in $X_1$, in agreement with the intrinsic degree of $\ell_3$ of $+1$. 

Above we have given only the first non-trivial steps in the construction of an $L_{\infty}$ 
algebra (that, however, captures already the relevant case of a Courant algebroid to be discussed 
shortly). In subsequent sections, the need for `higher' brackets and gauge structures will, however, 
reemerge in the construction of tensor hierarchies, where the higher $L_{\infty}$ brackets are (partly) 
encoded in higher-form field strengths, Chern--Simons terms, etc.

\section{Generalized diffeomorphisms}\label{sec:3}

\subsection{General construction}

We will now show that the embedding tensor formalism introduced above can be 
used to derive the generalized diffeomorphisms of double and exceptional field theory 
from an infinite-dimensional Lie algebra. This Lie algebra is an extension of the 
diffeomorphism algebra (in typically very large dimensions) by a `current algebra' 
based on a U-duality algebra. 
Specifically, let $\mathfrak{g}_0$ be the Lie algebra of a U-duality group such as E$_{6(6)}$, E$_{7(7)}$ or E$_{8(8)}$ 
with generators $t_{\alpha}$ satisfying  
 \be
  [t_{\alpha}, t_{\beta}] \ = \ f_{\alpha\beta}{}^{\gamma}t_{\gamma}\;.
 \ee  
Moreover, we have to pick a representation space $R$ of $\mathfrak{g}_0$
with representation matrices $(t_{\alpha})_M{}^{N}$, where $M, N=1,\ldots,{\rm dim}(R)$. 
For the exceptional field theories this is the representation in which vector fields transform in.
The infinite-dimensional Lie algebra ${\mathfrak{G}}$ is now defined by introducing 
coordinates $Y^M$ for this space, 
which for now we can take to be $\mathbbm{R}^{{\rm dim}(R)}$, 
and defining functions of these coordinates. 
These functions are given by pairs $\zeta  \equiv  (\lambda^M, \sigma^{\alpha})$,  
and the Lie brackets read 
 \be\label{LieBracketss}
 \begin{split}
  [\zeta_1,\zeta_2] \ = \ \big(2\,&\lambda_{[1}{}^{N}\partial_N\lambda_{2]}{}^{M} ,\, \\
  & 2\,\lambda_{[1}{}^{N}\partial_N\sigma_{2]}{}^{\alpha}
   +    f_{\beta\gamma}{}^{\alpha} \sigma_1{}^{\beta}\sigma_{2}{}^{\gamma}\big) \;. 
 \end{split}  
 \ee
In the first component this is the familiar diffeomorphism algebra for vector fields 
$\lambda^M$. The second component indicates that the $\sigma^{\alpha}$ 
are scalars 
under these diffeomorphisms  
and  live in the 
adjoint representation of the original Lie algebra $\frak{g}_0$. 
Since the diffeomorphism algebra is a Lie algebra, and since the action on $\sigma^{\alpha}$ is a representation, 
the bracket (\ref{LieBracketss}) defines a genuine Lie algebra obeying  the Jacobi identity. 
In particular, the dependence of 
$\zeta  =  (\lambda^M, \sigma^{\alpha})$ on the ${\rm dim}(R)$ coordinates is completely general 
(up to reasonable smoothness assumptions) and not constrained by any `section conditions'.

Next, we discuss some representations of the infinite-dimensional Lie algebra $\frak{G}$
defined by (\ref{LieBracketss}). 
The adjoint representation acts on $a =(p^M, q^{\alpha})\in \frak{G}$ as $\delta_{\zeta}a  =  [\zeta, a]$, 
which yields in components 
 \be
  \begin{split}
   \delta_{\zeta}p^M \ &= \ \lambda^N\partial_Np^M - p^N\partial_N\lambda^M\;, \\
   \delta_{\zeta}q^{\alpha} \ &= \ \lambda^N\partial_N q^{\alpha}-p^N\partial_N\sigma^{\alpha} 
   +f_{\beta\gamma}{}^{\alpha} \sigma^{\beta} q^{\gamma}\;. 
  \end{split}
 \ee
The coadjoint representation acts on $\frak{G}^*$, whose elements are  functions 
${\cal A}  = (A_{\alpha}, B_M)$, for which the pairing $\frak{G}^*\otimes \frak{G}\rightarrow \mathbbm{R}$ 
is given by the integral: 
 \be\label{integralPAIRING}
  {\cal A}(a) \ = \ \int {\rm d}Y\big(p^M B_M \ + \ q^{\alpha}A_{\alpha}\big) \;, 
 \ee
where ${\rm d}Y\equiv {\rm d}^{{\rm dim}(R)}Y$.  
The coadjoint action is determined by requiring invariance of this integral: 
 \be\label{coadjointaction}
 \begin{split}
  \delta_{\zeta} A_{\alpha} \ = \ &\,\lambda^N\partial_N A_{\alpha}+\partial_N\lambda^N A_{\alpha} 
  +f_{\alpha\beta}{}^{\gamma} \sigma^{\beta} A_{\gamma}\;, \\
  \delta_{\zeta} B_M \ = \ &\, \lambda^N\partial_N B_M + \partial_M\lambda^N B_N +\partial_N\lambda^N B_M\,+\\
  & +A_{\alpha}\,\partial_M\sigma^{\alpha}\;.  
 \end{split}
 \ee

The representation 
$R$ extends to a representation on the vector space of 
$R$-valued functions $V^M(Y)$, where the action of $\zeta\in \mathfrak{g}$ is given by 
 \be\label{bigaction}
  \delta_{\zeta}V^M \ \equiv \
   \lambda^N\partial_NV^M +  \gamma \, \partial_N\lambda^N V^M 
  - \sigma^{\alpha}(t_{\alpha})_N{}^{M}V^N\;.  
 \ee
Here we have included an arbitrary density weight $\gamma$, and sometimes we denote the 
representation space as $R^{[\gamma]}$. 
Using that the $(t_{\alpha})_M{}^{N}$ form a representation of the original algebra $\frak{g}_0$, 
it is straightforward to verify that (\ref{bigaction}) is indeed a representation of (\ref{LieBracketss}). 
More generally, we can canonically define representations on any 
tensor power of $R$. In addition, there is the dual representation $R^*$, whose elements 
are functions $A_M$ with invariant pairing $R\otimes R^*\rightarrow \mathbbm{R}$ given by 
 \be
  A(V) \ \equiv \ \int {\rm d}Y\, V^M A_M\;. 
 \ee  
Upon requiring invariance of this integral one infers that the dual space $(R^{[\gamma]})^*$ to the representation space $R^{[\gamma]}$ 
consists of functions $A_M$ of intrinsic density weight $1-\gamma$, with the transformation rules 
  \be\label{starRaction}
  \begin{split}
  \delta_{\zeta}A_M
   \ \equiv \ &\,  \lambda^N\partial_NA_M 
  + (1-\gamma)   \partial_N\lambda^N A_M\\
  &\, +\sigma^{\alpha} (t_{\alpha})_{M}{}^{N}A_N\;. 
 \end{split} 
 \ee 

In the following, we will have to refine this structure in order to
define a consistent algebra of generalized diffeomorphisms.
First of all, the coordinate dependence of the functions $\zeta$  
--- and more generally of all the function spaces we will be working with --- 
will be restricted. 
There will therefore be non-vanishing vectors $\lambda^M$ so that $\lambda^M\partial_M=0$ 
acting on any functions belonging to the same class,
in particular the $\zeta$ in (\ref{LieBracketss}).
As a consequence, the subalgebra ${\frak I}$ defined as
\bea
{\frak G} \supset {\frak I} = \left\{ \,\zeta=(\lambda^M,0)\in\frak{G} \big|  \lambda^M\partial_M  = 0 \,\right\}
\;,
\label{ideal}
\eea
is generally non-empty, 
forming an Abelian ideal of $\frak{G}$.
The subsequent construction is based on the coset algebra
 \bea
 \frak{g} &=& \frak{G}/\frak{I}
 \;.
 \label{coset}
 \eea
 Its dual $\frak{g}^*$ is made from elements ${\cal A}=(A_\alpha, B_M)$
 via a pairing (\ref{integralPAIRING}), where the non-trivial denominator of (\ref{coset})
 requires the functions $B_M$ to satisfy
 \bea
 \forall \lambda^M : \;\;\lambda^M \partial_M = 0\;\;\Longrightarrow\;\; \lambda^M B_M = 0
 \;.
 \label{ccc}
 \eea
 More generally, any $\frak{G}$ representations discussed above immediately lift to 
 representations of the coset algebra $\frak{g}$, assuming that the corresponding functions are subject 
 to the same restrictions. 
 In exceptional field theories, the representation $\frak{g}^*$ is typically assigned to the
 $(n-2)$ forms, $n$ referring to the number of external dimensions. 
 This has its origin in the fact that conserved currents associated to
global symmetries $\frak{g}$ may be dualized into Abelian forms of this rank. 
Indeed, these forms appear in pairs $(A_\alpha, B_M)$, with the second
component restricted by (\ref{ccc}). These were originally found as `covariantly constrained' compensator gauge fields
and required for a proper description of the dual graviton degrees of freedom~\cite{Hohm:2014fxa,Hohm:2018qhd}.

In the remainder of this subsection we discuss the specific structure of the embedding tensor map  
$\vartheta: R \rightarrow  \frak{g}$ for the above representation $R$. 
Given $V^M\in R^{[\gamma]}$, a natural ansatz is 
 \be\label{VARtheta}
  \vartheta(V) \ = \ \big[\big(V^M\,,\;-\kappa (t^{\alpha})_{M}{}^{N}\partial_N V^M\big) \big] \ \in \ \frak{g}\;, 
 \ee 
where $\kappa$ is a parameter that in examples is fixed by the quadratic constraint, and 
$[\,]$ indicates the equivalence class identifying two functions whose difference lies in the ideal (\ref{ideal}). 
From this we can compute the form of the embedding tensor defined by $\Theta(V,{\cal A}) = -{\cal A}(\vartheta(V))$, 
c.f.~eq.~(\ref{THETAbig}). Using the pairing (\ref{integralPAIRING}), one finds 
for $V\in {R}$, ${\cal A}=(A_{\alpha}, B_M)\in \frak{g}^*$ 
 \be\label{genTheta}
  \Theta(V, {\cal A}) \ = \
   - \int {\rm d}Y\big(V^M B_M \ - \  \kappa\, (t^{\alpha})_M{}^{N} A_{\alpha}\,\partial_NV^M\big)\;. 
 \ee
Below we will verify the quadratic constraint by proving invariance of this integral. 

We can now define the generalized Lie derivatives w.r.t.~$\Lambda\in R$ as the Leibniz action (\ref{GENLeibnizZZ}), 
 \be\label{generalizedLIE}
  {\cal L}_{\Lambda}V^M \ \equiv \ \delta_{\vartheta(\Lambda)}V^M\;, 
 \ee
where the right-hand side denotes the representation (\ref{bigaction}).  
Then, using (\ref{VARtheta}) in (\ref{bigaction}) we obtain 
 \be\label{generalgeneralLIE}
 \begin{split}
  {\cal L}_{\Lambda}V^M \ \equiv \ &\,\Lambda^N\partial_N V^M  \ + \ \gamma\,\partial_N\Lambda^N V^M\,+ \\
  &\,+  \ \kappa\,(t^{\alpha})_N{}^M(t_{\alpha})_L{}^{K}\,
  \partial_K\Lambda^L\,V^N
  \;, 
 \end{split}
 \ee 
which is the general form of the generalized Lie derivative in double and exceptional field theory \cite{Berman:2011jh}. 
As will be established below, the quadratic constraints and hence closure of the generalized Lie derivatives 
requires `section constraints' of the form 
\bea
 Y^{MN}{}_{KL}\,\partial_M \otimes \partial_N &=& 0
 \;,
\label{section0}
\eea
where $Y^{MN}{}_{KL}$ is a specific $\frak{g}_0$-invariant tensor,  
and the notation indicates an action of the differential operators on any pair of functions. 
We then infer that the ideal (\ref{ideal}) contains elements of the form $(Y^{MN}{}_{KL}\,\partial_N \chi^{KL},0)$. 
Moreover, the general discussion of section~\ref{sec:higher} has revealed the existence of trivial gauge parameters,
i.e.\ of a non-vanishing kernel of $\vartheta$. Specifically, following the discussion after (\ref{DREL}), 
this kernel contains the image of the ${\cal D}$ operator:
\bea
\vartheta \circ {\cal D}=0
\;.
\label{thD0}
\eea
We finally note that 
any representation of $\frak{g}$, such as the adjoint and coadjoint representation,  
can be lifted to a representation of the Leibniz algebra on $R$ by taking the infinitesimal parameter 
to be $\vartheta(\Lambda)$. 
We have thus obtained the generalized Lie derivatives from a Lie algebra and an embedding tensor, 
but it remains to verify the quadratic constraint. So far this can only be done on a case-by-case basis, 
to which we turn in the next subsections.

\subsection{$O(d,d)$ generalized diffeomorphisms}

We first consider the T-duality group $O(d,d)$ that is relevant for double field theory. 
The representation $R$ is given by the $2d$-dimensional fundamental representation, 
with fundamental indices $M, N=1,\ldots, 2d$. 
The structure constants 
and representation matrices are given by 
 \be\label{structureconst}
 \begin{split}
  f^{IJ,KL}{}_{PQ} \ &= \ 8\,\delta^{[I}{}_{[P}\,\eta^{J][K}\delta^{L]}{}_{Q]}\;,\\
  (t^{IJ})_M{}^{N} \ &= \ 2\,\delta^{[I}{}_{M}\,\eta^{J]N}\;, 
 \end{split}
 \ee
with the  $O(d,d)$ invariant metric 
\be
\label{etais}
\eta_{MN} \ = \  \begin{pmatrix}
0&{\bf 1} \\{\bf 1}&0 \end{pmatrix} \;, 
\ee
where ${\bf 1}$ denotes the $d\times d$ unit matrix.
The adjoint index is given by index pairs, $\alpha=[IJ]$, and we follow the convention that summation 
over such index pairs is accompanied by a factor $\frac{1}{2}$.

The infinite-dimensional Lie algebra $\frak{g}$ described above then consists of functions  $\zeta=(\lambda^M, \sigma^{IJ})$. 
The embedding tensor map $\vartheta :\,R\rightarrow\frak{g}$ in (\ref{VARtheta}) reduces to 
 \be\label{OddEmbedding}
 \begin{split}
  \vartheta(V) \ &= \ 
  \big(V^M\,, \;-\kappa(t^{IJ})_M{}^{N}\partial_NV^M)\\
  \ &= \ \big(V^M\,,\; K^{IJ}(V)\big)\;, 
 \end{split}
 \ee
where we set $\kappa=1$, which will be confirmed below. Moreover, we defined 
 \be\label{KIJ}
  K_{IJ}(V) \ \equiv \ \partial_{I} V_{J} - \partial_{J} V_{I}\;, 
 \ee
where here and in the following $O(d,d)$ indices will be raised and lowered with the metric (\ref{etais}). 
Similarly, the integral form (\ref{genTheta}) of the embedding tensor
for the coadjoint vector 
 ${\cal A}  \equiv (A_{IJ}, B_M)$
reduces to 
 \be\label{Oddembedding}
 \begin{split}
  \Theta(V,{\cal A}) \ &= \ -\int {\rm d}Y\big(V^M B_M \ - \ \tfrac{1}{2}(t^{IJ})_M{}^{N} A_{IJ}\,\partial_NV^M\big) \\
  \ &= \  -\int {\rm d}Y\big(V^M B_M \ + \ \tfrac{1}{2}A_{IJ}K^{IJ}(V)\big)\;. 
 \end{split}
 \ee

Before turning to the discussion of quadratic constraints, let us spell out 
the action of the generalized Lie derivative on various tensors. 
First, for a fundamental vector $V^M$ (of intrinsic density weight $\gamma$), eq.~(\ref{bigaction})
yields 
 \be\label{OddLieACTION}
  \delta_{\zeta}V^M \ = \ \lambda^N\partial_NV^M + \gamma\, \partial_N\lambda^N V^M - \tfrac{1}{2}\sigma^{IJ} (t_{IJ})_N{}^{M} V^N\;. 
 \ee
Together with (\ref{OddEmbedding}) we can then determine the generalized Lie derivative (\ref{generalizedLIE}), 
 \be\label{gammaderivative}
  {\cal L}_{\Lambda}^{[\gamma]}V^M \ = \ \Lambda^N\partial_N V^M +\gamma\, \partial_N\Lambda^N V^M +  K^{M}{}_{N}(\Lambda) V^N\;. 
 \ee  
Here and in the following we sometimes indicate the density weight by a superscript on ${\cal L}$. 
This notion of generalized Lie derivatives straightforwardly extends to arbitrary tensors of $O(d,d)$, 
where each index is accompanied by a `rotation term' employing the matrix $K$.   
For a tensor $T^M{}_N$ of density weight zero we have 
 \be
  {\cal L}_{\Lambda}^{[0]}T^M{}_{N} \ = \ \Lambda^K\partial_K T^M{}_{N} + K^{M}{}_{K} T^{K}{}_{N}+ K_{N}{}^{K} T^{M}{}_{K}\;, 
 \ee
and similar formulas readily follow for higher tensors. 

While in (\ref{gammaderivative}) we have given the generalized Lie derivative 
for arbitrary density weight $\gamma$, we will see in a moment that invariance of (\ref{Oddembedding}) 
requires $\gamma=0$ for $V^M$. Thus, the Leibniz algebra (here also referred to as  Dorfman bracket) is defined on the space of weight-zero vectors by 
$V\circ W={\cal L}_{V}^{[0]}W$, i.e., 
 \be\label{OddLeibniz}
  (V\circ W)^M \ = \ V^N\partial_NW^M + \partial^MV_N W^N -\partial_NV^M W^N\;. 
 \ee
From this one infers the symmetric part 
 \be
  \{V,W\}^M \ = \ \tfrac{1}{2}\partial^M\big(V_N W^N\big)\;, 
 \ee
and the antisymmetric part (the so-called `C-bracket') 
 \be\label{CBracket}
 \begin{split}
  [V,W]^M \ = \ &\, V^N\partial_N W^M-W^N\partial_N V^M\\
  &\, - \tfrac{1}{2} V_N\,\partial^M W^N +\tfrac{1}{2} W_N\,\partial^M V^N\;, 
 \end{split}
 \ee
where we used the notation (\ref{symmantisymm}). In particular, we see that $\{V,W\}=\frac{1}{2}{\cal D}(V\bullet W)$, c.f.~(\ref{DREL}), holds, 
using the notation of sec.~\ref{Linfsection}, for  
 \be\label{DREAL1}
  {\cal D}\,:\;X_1\rightarrow X_0\;, \quad ({\cal D}f)^M \ \equiv \ \partial^Mf\;, 
 \ee
and 
 \be\label{DREAL2}
  \bullet\,:\;X_0\otimes X_0\rightarrow X_1\;, \quad V\bullet W \ \equiv \ \eta_{MN} V^M W^N\;, 
 \ee
where $X_0$ is the space of $O(d,d)$ vectors  
and $X_1$ the space of $O(d,d)$ scalars. 

We now turn to the section constraint, which restricts the coordinate dependence of all functions and is needed for  the consistency of the above construction. 
In fact, we will see that the quadratic constraint is not satisfied unless such a constraint is imposed. 
For the $O(d,d)$ case, this constraint (which originates from the level-matching constraint of string theory and is sometimes referred to as the weak constraint) takes the form 
 \be\label{weakconstraint}
  \eta^{IJ}\partial_I\partial_J f \ \equiv \ \partial^I\partial_If \ = \ 0\;, 
 \ee
for any functions $f$. Splitting up the $2d$ coordinates into $d$ `momentum' coordinates and $d$ `winding' coordinates, 
this constraint is solved, owing to the split signature of (\ref{etais}),  by functions depending only on one set of coordinates 
(although in this weak form there are more general solutions, with functions depending both on momentum and winding coordinates).  
However, since the differential operator entering (\ref{weakconstraint}) is second-order, a subtle consistency issue arises: 
the product of two functions satisfying (\ref{weakconstraint}) does not necessarily satisfy the same constraint since 
$\partial^I\partial_I(f_1\cdot f_2)=2\,\partial^If_1\,\partial_If_2$. 
For now we circumvent this issue by simply \textit{demanding} the functions to be closed under multiplication, 
which amounts to imposing 
 \be\label{strongconstraint}
  \partial^If_1\,\partial_If_2 \ = \ 0\;, 
 \ee
for any functions $f_1$, $f_2$. 
This is the $O(d,d)$ version of (\ref{section0}) with $Y^{MN}{}_{KL}=\eta^{MN}\eta_{KL}$\,.
Together, (\ref{weakconstraint}) and (\ref{strongconstraint}) are referred to as the 
strong section constraint. One can then show that the most general solution of the strong constraint is given by 
functions depending only on $d$ coordinates. 

Accordingly, the ideal (\ref{ideal}) within $\frak{G}$ is non-empty, and the 
coadjoint representation $\frak{g}^*$ is spanned by vectors ${\cal A}  \equiv (A_{IJ}, B_M)$
of which the latter functions are constrained according to (\ref{ccc}) to satisfy
 \be\label{Bconstraint}
  B_M\,\partial^M \ = \ \partial^M B_M \ = \ 0\;. 
 \ee
Next, we compute the transformation of such a coadjoint 
vector 
 w.r.t.~$\Lambda\in R$. 
From the first equation in (\ref{coadjointaction}) we obtain 
 \be
 \begin{split}
  \delta_{\Lambda}A_{IJ} \ = \ &\, \vartheta(\Lambda)^N\partial_N A_{IJ} + \partial_N\vartheta(\Lambda)^N A_{IJ}\,+\\
  &\, +\tfrac{1}{4}\, f_{IJ,KL}{}^{PQ}\vartheta(\Lambda)^{KL} A_{PQ} \\
  \ = \ &\, \Lambda^N\partial_N A_{IJ} +\partial_N\Lambda^N A_{IJ} +2\,K_{[I}{}^{K}(\Lambda) A_{|K|J]}\;, 
 \end{split}
 \ee
using (\ref{structureconst}) and (\ref{OddEmbedding}).  
We observe that this takes the form of a generalized Lie derivative of a second-rank antisymmetric tensor 
of density weight one. 
From the second equation in (\ref{coadjointaction}) we obtain similarly 
 \be
 \begin{split}
  \delta_{\Lambda}B_M \ = \ &\, \Lambda^N\partial_N B_M +\partial_M\Lambda^N B_N + \partial_N\Lambda^N B_M \\
  &\, +\tfrac{1}{2}\,\partial_MK^{IJ}(\Lambda) A_{IJ}\;. 
 \end{split}
 \ee 
The first line can be rewritten as a generalized Lie derivative of a vector of weight one, using that one 
term in $K(\Lambda)_M{}^N B_N$ vanishes due to (\ref{Bconstraint}). 
Summarizing our results for the fields entering the integral (\ref{Oddembedding}), we have  
 \be\label{totalvariationsss}
  \begin{split}
   \delta_{\Lambda}V^M \ &= \ {\cal L}_{\Lambda}^{[0]}V^M\;, \\
   \delta_{\Lambda} A_{IJ} \ &= \ {\cal L}_{\Lambda}^{[1]}A_{IJ}\;, \\
   \delta_{\Lambda}B_M \ &= \ {\cal L}_{\Lambda}^{[1]}B_M +\tfrac{1}{2}\,\partial_MK^{IJ}(\Lambda) A_{IJ}\;. 
  \end{split}
 \ee
 
Our goal is now to prove invariance of $\Theta$.  
In order to compute the variation of $\Theta$ efficiently, we introduce a notation for `non-covariant' 
variations, the difference between the actual variation and the covariant one given by the (generalized) 
Lie derivative, 
 \be\label{DELTALAMBDA}
  \Delta_{\Lambda} \ \equiv \ \delta_{\Lambda}-{\cal L}_{\Lambda}\;. 
 \ee
For instance, (\ref{totalvariationsss}) is then expressed as $\Delta_{\Lambda}V^M=\Delta_{\Lambda} A_{IJ}=0$
and 
 \be\label{DeltaBBBBB}
  \Delta_{\Lambda}B_M \ = \ \tfrac{1}{2}\,\partial_MK^{IJ}(\Lambda) A_{IJ}\;. 
 \ee
We next compute the variation of 
the tensor (\ref{KIJ}), 
 \be\label{deltaKK}
 \begin{split}
  \delta_{\Lambda}K_{IJ}(V) =  &\,2\,\partial_{[I}\big(\Lambda^K\partial_K V_{J]} + K_{J]}{}^{K}(\Lambda)V_{K}\big) \\
   =  &\,\Lambda^K\partial_K K_{IJ}(V)  + 2\,(\partial_{[I}\Lambda^K-\partial^K\Lambda_{[I}) \partial_K V_{J]}\,- \\
  &\,-2\,K_{[I}{}^{K}(\Lambda)\partial_{J]}V_K \ + \ 2\partial_{[I}K_{J]}{}^{K}(\Lambda) V_K\;, 
 \end{split} 
 \ee
where we added a term in the second line that is zero by the constraint (\ref{strongconstraint}). 
We recognize the generalized Lie derivative of $K_{IJ}(V)$, up to terms involving $\partial K(\Lambda)$. Upon simplifying 
the latter terms, one obtains 
 \be
  \delta_{\Lambda}K_{IJ}(V) \ = \ {\cal L}_{\Lambda}K_{IJ}(V) - \partial^K K_{IJ}(\Lambda) V_K\;, 
 \ee
and thus, in terms of (\ref{DELTALAMBDA}), 
 \be\label{DELTAKKKKKK}
  \Delta_{\Lambda}K_{IJ}(V) \ = \ -V^K\partial_KK_{IJ}(\Lambda)\;. 
 \ee 
It is now straightforward to prove invariance of (\ref{Oddembedding}) under (\ref{totalvariationsss}). 
We first note that both terms under the integral are $O(d,d)$ scalars of density weight 1, exactly as 
needed for invariance. Thus, it only remains to verify cancellation of  the non-covariant variations, 
which is immediate with (\ref{DeltaBBBBB}) and (\ref{DELTAKKKKKK}): 
 \be
 \begin{split}
  \Delta_{\Lambda}\Theta(V,{\cal A}) &=  -\int {\rm d}Y\big(V^M\Delta_{\Lambda}B_M + \tfrac{1}{2}A_{IJ} \Delta_{\Lambda}K^{IJ}(V)\big)\\
   &=  0\;.  
 \end{split}
 \ee
Thus, the quadratic constraints are satisfied, which implies that the generalized Lie derivatives 
define a Leibniz algebra and hence close.

Next, we display the quadratic constraint in the form (\ref{finalquadrconstr2346}), which is written 
in terms of the embedding tensor map $\vartheta$. 
We compute with (\ref{LieBracketss}) and (\ref{OddEmbedding}) 
 \be\label{thetaLieBracket}
 \begin{split}
  [\vartheta(\Lambda_1), \vartheta(\Lambda_2)] \ = \ \,\big(&2\,\Lambda_{[1}{}^{N}\partial_N\Lambda_{2]}{}^{M}\,,\; \\
  &\;\;2\,\Lambda_{[1}{}^{N}\partial_NK_{2]}{}^{IJ} +2\, K_{[1}{}^{I}{}_{K} K_{2]}{}^{KJ}\big)\;, 
 \end{split} 
 \ee
where we used the short-hand notation $K_1=K(\Lambda_1)$, etc. 
On the other hand, 
 \be
 \begin{split}
  \vartheta(\Lambda_1\circ \Lambda_2) \ = \ \big(&{\cal L}_{\Lambda_1}\Lambda_2{}^{M}\,,\; 
  K^{IJ}(\Lambda_1\circ \Lambda_2)\big) \\
  \ = \ \big(&2\,\Lambda_{[1}{}^{N}\partial_N\Lambda_{2]}{}^{M}
  +\partial^M\Lambda_{1N}\Lambda_{2}{}^{N}\,,\;\\
  &\;\; K^{IJ}(\Lambda_1 \circ \Lambda_2)\big)\;. 
 \end{split}
 \ee
Comparing with (\ref{thetaLieBracket}) we seem to infer a mismatch in the first argument by the term $\partial^M\Lambda_{1N}\Lambda_{2}{}^{N}$.
However, as this term vanishes upon contraction with $\partial_M$ by virtue of the section constraint (\ref{strongconstraint}), the
discrepancy lives within the ideal ${\frak I}$ of (\ref{ideal}), thus vanishes within the coset $\frak{g}$.

As a result, the quadratic constraint in the form (\ref{finalquadrconstr2346})
 implies that
 \be
  K^{IJ}(\Lambda_1 \circ \Lambda_2) \ = \ 2\,\Lambda_{[1}{}^{N}\partial_NK_{2]}{}^{IJ} +2\, K_{[1}{}^{I}{}_{K}\, K_{2]}{}^{KJ}\;, 
 \ee
which one may also verify by a direct computation.

The above treatment of generalized Lie derivatives, based on the general abstract theory of sec.~\ref{embeddingtensorsec}, 
allowed us to obtain all formulas characterizing the gauge structure of double field theory without any significant 
calculations (the only real computation being (\ref{deltaKK}) that proves (\ref{DELTAKKKKKK})). 
While in the standard formulation of double field theory the coadjoint fields ${\cal A}$ do not enter, 
we think that the above discussion is illuminating in that it outlines the universal role of these fields. 
Indeed, these fields play a much more prominent role for higher-rank exceptional groups, 
notably for the E$_{8(8)}$ theory to which we turn momentarily, where they are indispensable in order to write a Lagrangian.

\subsection{E$_{7(7)}$ generalized diffeomorphisms}\label{E7subsection}

Let us start by summarizing the relevant features of the exceptional Lie group E$_{7(7)}$, whose 
Lie algebra is of dimension 133, with generators $t_{\alpha}$, $\alpha=1,\ldots, 133$. 
The fundamental representation is 56-dimensional, with indices $M,N=1,\ldots,56$. The symplectic embedding
E$_{7(7)}\subset {\rm Sp}(56)$ yields an invariant antisymmetric tensor $\Omega^{MN}$, 
which we use to raise and lower fundamental indices:
$V^M=\Omega^{MN} V_N$, $V_M=V^N\Omega_{NM}$, where $\Omega^{MK}\Omega_{NK} = \delta_{N}{}^{M}$. 
Adjoint indices are raised and lowered by the (rescaled) symmetric Cartan--Killing form
$\kappa_{\alpha\beta}\equiv (t_{\alpha})_M{}^N (t_{\beta})_N{}^M$.
Due to the invariance of $\Omega^{MN}$, the gauge group generators with 
index structure $(t_{\alpha})_{MN}$ are symmetric.   
The projector onto the adjoint representation is given by 
\bea\label{adjproj}
\mathbbm{P}^K{}_M{}^L{}_N&\equiv&
(t_\alpha)_M{}^K (t^\alpha)_N{}^L \nonumber\\ 
&=&
\frac1{24}\,\delta_M^K\delta_N^L
+\frac1{12}\,\delta_M^L\delta_N^K
+(t_\alpha)_{MN} (t^\alpha)^{KL}\,-\\ \nonumber
&& -\frac1{24} \,\Omega_{MN} \Omega^{KL}
\;. 
\eea
The generalized Lie derivative (\ref{generalgeneralLIE}) reads 
 \be\label{genLie}
 \begin{split}
{\cal L}^{[\gamma]}_{\Lambda} V^M 
\  \equiv \ \,&\Lambda^K \partial_K V^M 
+\gamma\,\partial_K \Lambda^K\,V^M\,-\\ 
& \quad 
- 12\, \mathbbm{P}^M{}_N{}^K{}_L\,\partial_K \Lambda^L\,V^N
\;,
\end{split}
\ee
where closure requires $\kappa=-12$ and the following section constraints: 
  \be
 \begin{split}
  (t_\alpha)^{MN}\,\partial_M \partial_N f \ &= \ 0\;, \quad   (t_\alpha)^{MN}\,\partial_Mf\, \partial_N g \ = \ 0 \,,\\
  \Omega^{MN}\,\partial_Mf\, \partial_N g \ &= \ 0 \,, 
 \end{split}
 \label{sectioncondition}
 \ee  
for arbitrary functions $f,g$. As a consequence of these constraints, 
there are trivial gauge parameters of the form 
\bea
\Lambda^M \ \equiv \ (t^\alpha)^{MN}\partial_N \chi_\alpha\;, \qquad 
\Lambda^M \ = \  \Omega^{MN}\chi_{N} 
\;, 
\label{trivial}
\eea
with a covariantly constrained $\chi_M$. The coadjoint action $\delta_{\Lambda}{\cal A}={\rm ad}^*_{\vartheta(\Lambda)}{\cal A}$ on ${\cal A}=(A_{\alpha}, B_M) \in \frak{g}^*$, c.f.~(\ref{coadjointaction}), yields 
 \be\label{coadjointE7}
  \begin{split}
   \delta_{\Lambda}A_{\alpha} \ &= \ {\cal L}_{\Lambda}^{[1]}A_{\alpha}\;, \\
   \delta_{\Lambda} B_{M} \ &= \ {\cal L}_{\Lambda}^{[\frac{1}{2}]}B_M + 12\, A_{\alpha}(t^{\alpha})_K{}^{L}\partial_M\partial_L\Lambda^K\;. 
  \end{split}
 \ee
The first line employs the natural action of the generalized Lie derivative on the field $A_{\alpha}$ 
in the adjoint of E$_{7(7)}$, as in the first line of (\ref{coadjointaction}).  
In order to verify the variation of $B_M$ one has to recall that this field is `covariantly constrained', 
i.e., subject to the same constraints as the derivatives in (\ref{sectioncondition}) so that $(t^{\alpha})^{MN}B_M\partial_N=0$, etc. 
It is then straightforward to verify 
 \be
   {\cal L}_{\Lambda}^{[\frac{1}{2}]}B_M  \ = \ \Lambda^N\partial_N B_M + \partial_M\Lambda^N B_N + \partial_N\Lambda^N B_M\;, 
 \ee
from which the second relation in (\ref{coadjointE7}) quickly follows.

Let us now turn to the Leibniz algebra, which is defined on the space of vectors $V^M$ 
of density weight $\gamma=\frac{1}{2}$. Using this and (\ref{adjproj}) one finds 
\be
 \begin{split}
 (V\circ W)^M
 \ = \ &
 V^N \partial_N W^M-W^N \partial_N V^M - \tfrac{1}{2}\partial^M V_N \,W^N\,-\\
& -12\,
  (t_\alpha)^{MN} (t^\alpha)_{KL}\,\partial_N V^K W^L 
  \;, 
 \end{split} 
 \ee
where we recall that indices are raised and lowered with $\Omega^{MN}$.   
Our goal is now to write its symmetric part, which is found to be 
 \be\label{E7symmetric}
 \begin{split}
  \big\{ V,W\big\}^M \ = \ &-6(t^\alpha)^{MN} \partial_N\big((t_{\alpha})_{KL}V^KW^L\big)\,+\\
   &+\frac{1}{4}\big(V_N\,\partial^MW^N
   +W_N\,\partial^MV^N\big)\;, 
 \end{split}
 \ee  
as in (\ref{DREL}), so that 
$\{V,W\} = \tfrac{1}{2}{\cal D}(V\bullet W)$. 
We first define the bullet operation 
 \be
   \bullet\, :\;\;X_0\ \otimes \ X_0 \; \rightarrow \; X_1\cong \frak{g}^*\;,  
 \ee
where $X_0$ is the space of the Leibniz algebra, and $X_1$
is the coadjoint representation space $\frak{g}^*$ with elements  ${\cal A}=(A_{\alpha}, B_M)$. 
The bullet operation is defined by 
 \be\label{E7bullet}
 \begin{split}
  V\bullet W \ \equiv \ \Big(&(t_{\alpha})_{KL} V^K W^L,\,\\
  &\;\;\; \tfrac{1}{2}\big(V_N\,\partial_MW^N+W_N\,\partial_MV^N\big)\Big) \ \in \ \frak{g}^*\;, 
 \end{split}
 \ee
where the free index of the second component is carried by a derivative and hence compatible 
with being `covariantly constrained'. We have to verify that the above right-hand side indeed transforms 
as required by (\ref{coadjointE7}). Here one needs for the second component that for $V\in X_0$
 \be
 \begin{split}
  \delta_{\Lambda}\big(\partial_MV^N\big) \ = \ &\,{\cal L}_{\Lambda}^{[-\frac{1}{2}]}\big(\partial_MV^N\big)\,-
  \\&\, 
  -12\,\mathbbm{P}^N{}_{K}{}^{P}{}_{Q}\,
  \partial_M\partial_P\Lambda^Q V^K\;. 
 \end{split}
 \ee

Next, we need to define the map ${\cal D}\,: X_1\cong \frak{g}^*\rightarrow X_0$, 
which acts on ${\cal A}=(A_{\alpha}, B_M)$ as
 \be\label{E7calD}
  \big({\cal D}{\cal A}\big)^M \ \equiv \ -12\Big((t^{\alpha})^{MN}\partial_NA_{\alpha} \ - \ \frac{1}{12} \Omega^{MN}B_N\Big)\;. 
 \ee
This combination appeared already in \cite{Hohm:2013uia}, c.f.~eq.~(2.24),\footnote{More precisely, comparison 
with that formula requires the identification $A_{\alpha}=W_{\alpha}$, $B_M=-\tfrac{1}{2}W_M$.} where it was shown to be covariant under 
generalized diffeomorphisms. With (\ref{E7bullet}) and (\ref{E7calD}) it is now immediate that the 
symmetric part (\ref{E7symmetric}) takes the form (\ref{DREL}). Finally, the above operator is also useful in order to write the embedding tensor $\Theta$ 
in (\ref{genTheta}) in terms of the symplectic invariant $\Omega(V,W)\equiv\int {\rm d}Y\,\Omega_{MN}V^MW^N$, which is gauge invariant 
for vectors of density weight $\frac{1}{2}$, as 
 \be\label{E7embedding}
  \Theta(V,{\cal A}) \ = \ \Omega(V,{\cal D}{\cal A})\;. 
 \ee
This form makes the gauge invariance of $\Theta$ and hence closure of the gauge algebra manifest.

\subsection{E$_{8(8)}$ generalized diffeomorphisms} \label{E8subsection}

The structure of generalized diffeomorphisms for $\frak{g}_0 = {\rm E}_{8(8)}$
brings about some particular features. To begin with, the representation $R$ of $\frak{g}_0$
underlying the definition of the algebra (\ref{LieBracketss}) 
in this case is the adjoint representation 
itself. Accordingly, the algebra $\frak{g}$ is defined in terms of
pairs $\zeta  \equiv  (\lambda^M, \sigma_M)$,  
with $M=1, \dots, 248$ labeling
the adjoint representation of E$_{8(8)}$, such that
co-adjoint vectors are functions ${\cal A}=(A^M,B_M)$.
The vector fields of the theory on the other hand do not
transform in $R$ but in its full extension to $\frak{g}^*$.
This is a characteristic of theories with $n=3$ external dimensions
and in line with the general discussion of $(n-2)$ forms 
following (\ref{ccc}) above.

As a result, the embedding tensor (\ref{embeddingtensormap}) 
is a map $\vartheta\,: \; \frak{g}^*\;\rightarrow\; \frak{g}$
and induces a bilinear form on the dual space, 
 \be\label{Theta_Form}
  \Theta\;:\quad \frak{g}^*\,\otimes\ \frak{g}^*\quad \rightarrow\quad \mathbbm{R}\;,
 \ee 
 which must be symmetric in order to admit the construction of invariant action functionals.
Specifically, for the ${\rm E}_{8(8)}$ ExFT, the action of  the embedding tensor map 
on an element $\frak{g}^* \ni \Upsilon = (\Lambda^M, \Sigma_M)$
reads
 \be\label{varthetaE8}
 \begin{split}
  \vartheta(\Upsilon) \ &= \ \big(\vartheta(\Upsilon)^M, \, \vartheta(\Upsilon)_M\big) \\
  \ &= \ \big(\Lambda^M, \; f_{M}{}^{N}{}_{K}\partial_N\Lambda^K + \Sigma_M\big)\;,
 \end{split}
 \ee
 with the structure constants $f_{MN}{}^K$ of ${\rm E}_{8(8)}$ and adjoint indices $M$
 raised and lowered with the Cartan--Killing form.
In turn, the induced bilinear form (\ref{ThetaINV}) on $\frak{g}^*$ is given by
 \bea
  \Theta({\cal A}_1,{\cal A}_2)  &=&  - {\cal A}_1(\vartheta({\cal A}_2)) 
  \nonumber\\
  &=& - \int {\rm d}Y\big(
  2A_{1}{}^M B_{2M} +A_2{}^M B_{1M}\,- 
  \nonumber\\
  &&{}\qquad\qquad
  - f^{M}{}_{NK}A_1{}^{N} \partial_MA_2{}^{K}\big)\;, 
 \label{varthetaE8abs}
 \eea 
and plays a central role in the construction of the invariant
action functional. 
 As previously discussed, the generalized Lie derivative is obtained via 
 (\ref{bigaction}), (\ref{VARtheta}) and 
accordingly depends on two gauge parameters $\Lambda^M$, $\Sigma_M$.
On an adjoint vector $V^M$ (of density weight $\gamma$), it acts as 
 \be\label{usualgenLie}
  {\cal L}^{[\gamma]}_{\Upsilon}V^M \ = \ \Lambda^N\partial_N V^M 
  +f^{M}{}_{NK} R^N V^K 
 +\gamma\, \partial_N\Lambda^N V^M\;, 
 %\nonumber
 \ee
with
 \be\label{RDEF}
  R^M \ \equiv \ f^{MN}{}_{K}\,\partial_N\Lambda^K +\Sigma^M\;. 
 \ee
The quadratic constraint requires section constraints
  \be\label{secconstr}
  \begin{split}
  \eta^{MN}\partial_M\otimes  \partial_N \ & = \ 0 \;, \\
  f^{MNK}\partial_N\otimes \partial_K \ &= \ 0\;, \\
  (\mathbbm{P}_{3875})_{MN}{}^{KL} \partial_K\otimes \partial_L \ &= \ 0
  \;,
  \end{split}
 \ee 
to be imposed on partial derivatives and also on the gauge parameter $\Sigma_M$\,.
Here, $(\mathbbm{P}_{3875})$ denotes the projector onto the ${\bf 3875}$ representation
of E$_{8(8)}$ within the symmetric tensor product
${\bf 248}\otimes_{\rm sym}{\bf 248}$.

 For completeness, we also state the
explicit form of the associated Leibniz product 
  \be\label{LeibnizE8Prod}
  \Upsilon_1 \circ  \Upsilon_2 \ \equiv \ \Big(\,{\cal L}_{\Upsilon_1}^{[1]}\Lambda_2{}^M \;,\; \,
  {\cal L}_{\Upsilon_1}^{[0]}\Sigma_{2M} \ + \ \Lambda_2{}^N\partial_M R_{N}(\Upsilon_1)\,\Big)\;.
 \ee
Its symmetric part takes the form
 \bea\label{trivialE8Leibniz}
   \{\Upsilon_1 ,  \Upsilon_2 \}
     & =& \ \Big(\,7(\mathbbm{P}_{3875})^{MK}{}_{NL}\, 
   \partial_K\big(\Lambda_1^N\Lambda_2^L\big) + \tfrac{1}{8}\, \partial^M\big(\Lambda_1^N \Lambda_{2N}\big)\,+
   \nonumber\\
   &&{}\;\;\;\;  + f^{MN}{}_{K}\, \Omega_{N}{}^{K}\;,\;\; \partial_M\Omega_N{}^{N} + \partial_N\Omega_{M}{}^{N}\;\Big)\;, 
  \eea
with
 \be
  \Omega_{M}{}^{N} \ = \ \Lambda_{(1}{}^{N}\Sigma_{2)M}
  -\tfrac{1}{2}\, f^{N}{}_{KL}\,\Lambda_{(1}{}^{K}\,\partial_M\Lambda_{2)}{}^L\;. 
 \ee
In analogy with (\ref{E7bullet}), (\ref{E7calD}), this may be disentangled as
$\{V,W\} = \tfrac{1}{2}{\cal D}(V\bullet W)$ into a map ${\cal D}: X_1 \rightarrow X_0$,
and a bullet structure
 \be
   \bullet\, :\;\;X_0\ \otimes \ X_0 \; \rightarrow \; X_1\;,  
 \ee
where $X_1$ in this case 
is spanned by fields in the ${\bf 1}\oplus{\bf 3875}$ of E$_{8(8)}$
together with fields of index structure $C_{M}{}^{N}$,
covariantly constrained in the first index.
Specifically,
with $X_1 \ni {\cal B}=(C,C_{(3875)}^{KL},C_M{}^N)$,
 the ${\cal D}$ map takes the form
\bea\label{DBE8}
{\cal D}{\cal B}\!&=&\!
\Big( 14\,(\mathbbm{P}_{3875}){}^{MN}{}_{KL}\partial_N C_{(3875)}^{KL}+\frac{1}{4}\partial^M  C
  +2f^{MN}{}_{K} C_{N}{}^{K},
  \nonumber\\
&&{}\qquad
2\, \partial_N C_M{}^{N}
   +2\,\partial_M C_{N}{}^{N}
   \Big)
   \;.
\eea

\section{Tensor hierarchy}\label{sec:4}

\subsection{Generalities and double field theory}

We will now define gauge theories based on the above higher algebraic structures. 
This in turn necessitates the appearance of higher-form gauge potentials entering 
in the form of a `tensor hierarchy' \cite{deWit:2008ta}. 

We begin with the $O(d,d)$ case, but present the formulas in a general form
likewise applicable to the exceptional field theories.
 $O(d,d)$ is relevant 
to bosonic string theory in $D=n+d$ dimensions, where $d$ internal dimensions 
are toroidal and hence doubled. The (internal) metric and B-field are then   
unified in terms of a generalized metric  
 \be\label{firstH}
  {\cal H}_{MN} \ = \  \begin{pmatrix}    g^{ij} & -g^{ik}b_{kj}\\[0.5ex]
  b_{ik}g^{kj} & g_{ij}-b_{ik}g^{kl}b_{lj}\end{pmatrix} \ \in \ O(d,d)\;, 
 \ee
which transforms under generalized Lie derivatives (\ref{gammaderivative}) as 
$\delta{\cal H}_{MN} ={\cal L}_{\Lambda}^{[0]}{\cal H}_{MN}$. 
The need for higher-form potentials arises as follows. The generalized internal 
diffeomorphisms are parameterized by $\Lambda(x,Y)$, which depend on the 
(doubled) \textit{internal} coordinates $Y^M$ but also on the \textit{external} 
coordinates $x^{\mu}$. Correspondingly,  the full action involves derivatives such as 
$\partial_{\mu}{\cal H}_{MN}$  that do not transform covariantly under these gauge transformations. 
The resolution is familiar from gauge theories: 
one introduces gauge fields $A_{\mu}$ and covariant derivatives. 
However, since the gauge structure does not define a Lie algebra, the naive Yang--Mills type 
field strength for $A_{\mu}$ is not gauge covariant. This can be remedied by introducing 
2-forms, which exhibits  the beginning of a tensor hierarchy. 
Indeed, for generic groups this procedure does not stop here but rather 
requires the introduction of 3-forms and higher forms. However, for $O(d,d)$ the tensor hierarchy 
ending with 2-forms is exact, which is hence a nice model to begin with. 

We start by introducing a gauge vector $A_{\mu}{}^{M}$ taking values in the Leibniz algebra 
and defining the covariant derivative w.r.t.~$x^{\mu}$
 \be\label{covDER}
   D_{\mu} \ \equiv \ \partial_{\mu} \ - \ {\cal L}_{A_{\mu}}\;, 
 \ee
where the generalized Lie derivative acts according to the representation of the field 
on which $D_{\mu}$ acts. 
For instance, for an $O(d,d)$ vector field $\Lambda^M$ of density-weight zero, we can write 
 \be
  D_{\mu}\Lambda \ = \ \partial_{\mu}\Lambda - A_{\mu}\circ \Lambda  \;, 
 \ee  
with the Leibniz product (\ref{OddLeibniz}). The gauge transformations for $A_{\mu}$ 
then take the familiar Yang--Mills form 
 \be\label{deltaLambdaA}
  \delta_{\Lambda}A_{\mu}{}^M  \ = \ D_{\mu}\Lambda^M\;. 
 \ee
The covariant derivatives (\ref{covDER}) indeed transform covariantly: 
on a generic tensor $V$ we have 
 \be
  \begin{split}
   \delta_{\Lambda}(D_{\mu}V) \ = \ &\,\delta_{\Lambda}(\partial_{\mu}V-{\cal L}_{A_{\mu}}V) \\
   \ = \ &\,\partial_{\mu}({\cal L}_{\Lambda}V)-{\cal L}_{\partial_{\mu}\Lambda -A_{\mu}\circ \Lambda}V-{\cal L}_{A_{\mu}}{\cal L}_{\Lambda}V\\
   \ = \ &\, {\cal L}_{\Lambda}(\partial_{\mu}V-{\cal L}_{A_{\mu}}V) \\
   &\quad+{\cal L}_{A_{\mu}\circ \Lambda}V - [{\cal L}_{A_{\mu}}, {\cal L}_{\Lambda}]V\\
   \ = \ &\, {\cal L}_{\Lambda}(D_{\mu}V)\;, 
  \end{split}
 \ee 
using  the algebra (\ref{gaugealgebra}) of generalized Lie derivatives.  
This works as for standard Yang--Mills theory, but next we encounter an 
important difference: the candidate field strength
 \be
  F_{\mu\nu} \ \equiv \ \partial_{\mu}A_{\nu}-\partial_{\nu}A_{\mu} -\big[A_{\mu},A_{\nu}\big]\;, 
 \ee
with bracket (\ref{CBracket}), is not gauge covariant. 
In order to discuss this efficiently, it is helpful to first compute the variation of $F_{\mu\nu}$ under general $\delta A_{\mu}$: 
 \be
 \begin{split}
  \delta F_{\mu\nu} \ &= \ 2 \big(\partial_{[\mu}\,\delta A_{\nu]}-\big[A_{[\mu},\,\delta A_{\nu]}\big]\big) \\
  \ &= \ 2\big(\partial_{[\mu}\,\delta A_{\nu]} - A_{[\mu}\circ \delta A_{\nu]} +\{A_{[\mu},\, \delta A_{\nu]}\}\big)\\
  \ &= \ 2\,D_{[\mu}\,\delta A_{\nu]} + {\cal D}(A_{[\mu}\bullet \delta A_{\nu]})\;, 
 \end{split}
 \ee
using (\ref{DREL}) in the last step. Restoring $O(d,d)$ indices and using (\ref{DREAL1}), (\ref{DREAL2}) 
this reads 
 \be\label{deltaF}
   \delta F_{\mu\nu}{}^M \ = \ 2\,D_{[\mu}\,\delta A_{\nu]}{}^M \ + \  \partial^M\big(A_{[\mu}{}^N \delta A_{\nu]N}\big)\;. 
 \ee
This is close to the familiar `Ricci identity' of gauge theories that is used to prove covariance of field strengths, 
but here we encounter an `anomaly' term that, however,  is `${\cal D}$ exact'. 
This suggest to define a modified curvature with additional 2-forms as:
 \be
  {\cal F}_{\mu\nu} \ \equiv \ \partial_{\mu}A_{\nu}-\partial_{\nu}A_{\mu} -\big[A_{\mu},A_{\nu}\big]  \ - \  {\cal D}B_{\mu\nu}\;. 
  \label{generalF}
 \ee
Specifically, for $O(d,d)$ this reads 
 \be\label{OddFIeldstength}
  {\cal F}_{\mu\nu}{}^M \ \equiv \ \partial_{\mu}A_{\nu}{}^M-\partial_{\nu}A_{\mu}{}^M -\big[A_{\mu},A_{\nu}\big]^{M} \ - \  \partial^MB_{\mu\nu}\;, 
 \ee
where $B_{\mu\nu}$ is a singlet 2-form. Using (\ref{deltaF}) we can then write 
 \be\label{genFVAR}
  \delta {\cal F}_{\mu\nu} \ = \ 2\,D_{[\mu}\,\delta A_{\nu]}{}^M \ - \ \partial^M\Delta B_{\mu\nu}\;, 
 \ee
where 
 \be\label{DELTAB}
  \Delta B_{\mu\nu} \  \equiv \ \delta B_{\mu\nu} - A_{[\mu}{}^N \delta A_{\nu]N}\;. 
 \ee
These relations can now be used to establish gauge covariance of ${\cal F}_{\mu\nu}$ under (\ref{deltaLambdaA}), 
provided we assign a suitable gauge transformation to $B_{\mu\nu}$: 
 \be
  \begin{split}
   \delta_{\Lambda}{\cal F}_{\mu\nu} \ &= \ \big[D_{\mu},D_{\nu}\big]\Lambda^M \ - \ \partial^M\Delta_{\Lambda}B_{\mu\nu}\\
   \ &= \ -({\cal F}_{\mu\nu}\circ \Lambda)^M \ - \ \partial^M\Delta_{\Lambda}B_{\mu\nu}\\
   \ &= \ (\Lambda\circ {\cal F}_{\mu\nu})^M-2\{{\cal F}_{\mu\nu}, \Lambda\}^M\ - \ \partial^M\Delta_{\Lambda}B_{\mu\nu}\\
    \ &= \ (\Lambda\circ {\cal F}_{\mu\nu})^M \ - \ \partial^M\big(\Delta_{\Lambda}B_{\mu\nu}+{\cal F}_{\mu\nu}{}^N \Lambda_N\big)\\
    \ &= \ {\cal L}_{\Lambda} {\cal F}_{\mu\nu}{}^{M}\;, 
  \end{split}
 \ee
where we used that the field strength satisfies 
 \be
  \big[D_{\mu}, D_{\nu}\big] \ = \ -{\cal L}_{{\cal F}_{\mu\nu}}\;, 
 \ee
and we set 
 \be
  \Delta_{\Lambda}B_{\mu\nu} \ = \ -\Lambda^M {\cal F}_{\mu\nu M}\;. 
 \ee 
Thus, the field strength transforms covariantly under the Yang--Mills-like gauge transformations, 
but due to the 2-form potential there is also a new, `higher' gauge invariance with 1-form parameter $\Lambda_{\mu}$, 
so that 1- and 2-form gauge potentials transform in total as 
 \be
  \begin{split}
   \delta_{\Lambda}A_{\mu}{}^M \ &= \ D_{\mu}\Lambda^M + \partial^M\Lambda_{\mu} \,, \\
   \Delta_{\Lambda} B_{\mu\nu} \ &= \ 2\,D_{[\mu}\Lambda_{\nu]}-\Lambda^M {\cal F}_{\mu\nu M}\;, 
  \end{split}
 \ee
where the second line is written in terms of (\ref{DELTAB}). Invariance of the field strength under 
these 1-form transformations follows with (\ref{genFVAR}) and a quick computation establishing 
that for a scalar ${\cal L}_{\Lambda}(\partial^MS)=\partial^M({\cal L}_{\Lambda}S)$, so that 
$D_{\mu}(\partial^M\Lambda_{\nu})=\partial^M(D_{\mu}\Lambda_{\nu})$.

Having introduced a 2-form gauge potential it is natural to try to define a field strength for it. 
Such a field strength indeed exists and can be written as 
 \be\label{Gfieldstrength00}
  {\cal H}_{\mu\nu\rho} \ = \ 3\Big(D_{[\mu}B_{\nu\rho]} + A_{[\mu}\bullet\big(\partial_{\nu}A_{\rho]}
  -\tfrac{1}{3}\big[A_{\nu},A_{\rho]}\big]\big) \Big)\;, 
 \ee
 or, in terms of more explicit $O(d,d)$ language, 
 \be\label{Gfieldstrength}
  {\cal H}_{\mu\nu\rho} \ = \ 3\Big(D_{[\mu}B_{\nu\rho]} + A_{[\mu}{}^{N}\partial_{\nu}A_{\rho]N}
  -\tfrac{1}{3}A_{[\mu N}\big[A_{\nu},A_{\rho]}\big]^N_{ } \Big)\;. 
 \ee
In this case, the 3-form field strength is already fully gauge covariant, 
 $\delta_{\Lambda} {\cal H}_{\mu\nu\rho}  =  \Lambda^M\partial_M {\cal H}_{\mu\nu\rho}$. 
Moreover, we have the `hierarchical' Bianchi identities 
 \be
 \begin{split}
   3 D_{[\mu}{\cal F}_{\nu\rho]}{}^{M} \ +\ \partial^M {\cal H}_{\mu\nu\rho} \ &= \ 0\;, \\
   4\,D_{[\mu}\, {\cal H}_{\nu\rho\sigma]} \ - \ 3\,{\cal F}_{[\mu\nu}{}^M\,{\cal F}_{\rho\sigma]}{}_M
   \ &= \ 0\;,
  \end{split}   
\ee
as can be checked by an explicit calculation.  A special feature of the field strength (\ref{Gfieldstrength00}) 
is that it is gauge covariant without the need to introduce any 3-form gauge potentials. 
This is directly related to the fact that the underlying bracket (\ref{CBracket}), which reduces to the Courant bracket 
upon eliminating the winding coordinates, yields an $L_{\infty}$-algebra with no higher brackets than a 3-bracket \cite{Roytenberg:1998vn}.  
In other words, the Courant bracket yields a so-called `2-term' $L_{\infty}$-algebra, which is defined on the short complex (\ref{SIMPLEexactsequence}).
For more general (U-duality) groups this will not be the case, so that higher brackets and higher $p$-forms need to be 
introduced. 
 (We refer to sec.~3 of \cite{Hohm:2015xna} for a detailed discussion of the proof of gauge invariance 
and the Bianchi identities 
in this more general setting.)

We are now ready to display the complete double field theory action in a `split formulation' with 
$n$ external and $2d$ internal (doubled) coordinates. 
The fundamental fields are 
 \be
\left\{\;  g_{\mu\nu}\,, \; B_{\mu\nu}\,, \;  \phi\,, \;  {\cal H}_{MN}\,, \;  A_{\mu}{}^{M}\;\right\} \;, 
 \ee
which all depend on coordinates $(x^{\mu}, Y^M)$. The gauge transformations of the 1- and 2-forms have 
been discussed above. 
The (internal) generalized metric transforms w.r.t.~the generalized Lie derivative, and 
the $O(d,d)$ singlet fields $g_{\mu\nu}$ and $\phi$ 
transform as scalar densities of appropriate weights under generalized diffeomorphisms w.r.t.~$\Lambda^M$. 
The action is given by 
 \be\label{finalActionIntro}
  \begin{split}
   S \ &= \  \int {\rm d}x\, {\rm d}Y \sqrt{g}\,e^{-2\phi}\Big(  \widehat{R}
   +4D^{\mu}\phi D_{\mu}\phi -\tfrac{1}{12}{\cal H}^{\mu\nu\rho}{\cal H}_{\mu\nu\rho}\,+\\
   &+\tfrac{1}{8}D^{\mu}{\cal H}^{MN}D_{\mu}{\cal H}_{MN}
   -\tfrac{1}{4}{\cal H}_{MN}{\cal F}^{\mu\nu M}{\cal F}_{\mu\nu}{}^{N}-V\Big)\;,  
  \end{split}
 \ee  
with the potential (characterized by carrying  only internal derivatives $\partial_M$) 
 \be
  V(\phi,{\cal H},g) \ = \ -{\cal R}(\phi,{\cal H}) - \frac{1}{4}{\cal H}^{MN}\partial_Mg^{\mu\nu}\,\partial_N g_{\mu\nu} \;, 
 \ee 
where ${\cal R}$ is the scalar curvature of double field theory \cite{Hohm:2010pp}. 
Moreover, $\widehat{R}$ is the suitably covariantized external Ricci scalar. 
Upon setting $\partial_M=0$, (\ref{finalActionIntro}) reduces to the action computed by Maharana 
and Schwarz by dimensional reduction of the familiar low-energy action of string theory \cite{Maharana:1992my}. 
Thus, as outlined in 
the introduction, the above action provides the proper non-Abelian extension of that theory. 
Upon breaking the $O(d,d)$ symmetry by letting fields depend on $d$ coordinates among the $Y^M$, 
the theory is fully equivalent to the standard NS-NS action (\ref{NSNSaction}). 

Let us note that the action (\ref{finalActionIntro}) is manifestly invariant under (generalized) internal diffeomorphisms, 
but it also has a non-manifest invariance under external diffeomorphisms with parameters 
$\xi^{\mu}(x,Y)$. More precisely, the invariance under $Y$-\textit{independent} $\xi^{\mu}$ 
transformations is manifest since (\ref{finalActionIntro}) is covariant according to the usual 
tensor calculus. However, whenever $\partial_M\xi^{\mu}\neq 0$, all terms in the above action 
are linked under external diffeomorphisms, which indeed fixes all relative coefficients. 
This invariance cannot be made manifest (at least as far as we know) without re-introducing 
dual $\tilde{x}_{\mu}$ coordinates and elevating the action to a full-fledged double field theory 
based on $O(d+n, d+n)$. This was indeed the method by which (\ref{finalActionIntro}) was 
originally derived \cite{Hohm:2013nja}.

\subsection{E$_{7(7)}$ exceptional field theory}

We now discuss the E$_{7(7)}$ exceptional field theory, in parallel to the discussion of $O(d,d)$, 
starting with the tensor hierarchy. 
Specifically, using the notation of sec.~\ref{E7subsection} we can write the gauge transformations 
in the same universal form as for $O(d,d)$: for $A_{\mu},\Lambda \in X_0$ 
and $\Xi_{\mu}\in X_1\cong \frak{g}^*$ we have 
 \be
  \delta A_{\mu} \ = \ D_{\mu}\Lambda \ + \ {\cal D}\Xi_{\mu}\;,  
 \ee
where $\Xi_{\mu}=(\Xi_{\mu \alpha}, \Xi_{\mu M})$, with the second component being covariantly constrained. 
Similarly, for the 2-forms ${\cal B}_{\mu\nu}=(B_{\mu\nu \alpha}, B_{\mu\nu M}) \in \frak{g}^*$ the covariant variations read 
 \be
  \Delta {\cal B}_{\mu\nu} \ \equiv \ \delta {\cal B}_{\mu\nu} \  + \ A_{[\mu}\bullet \delta A_{\nu]}\;. 
 \ee
Using (\ref{E7bullet}) this can be written out as two relations, 
\bea
\Delta B_{\mu\nu\,\alpha} &\equiv& \delta B_{\mu\nu\,\alpha} +  
(t_\alpha)_{KL}\,A_{[\mu}{}^K\, \delta A_{\nu]}{}^L\;,
\\
\Delta B_{\mu\nu\,K} &\equiv& \delta B_{\mu\nu\,K}\,-\nonumber \\
&&-\tfrac{1}{2}\Omega_{NL}\left(A_{[\mu}{}^N \partial_K \delta A_{\nu]}{}^L
-\partial_K A_{[\mu}{}^N \, \delta A_{\nu]}{}^L\right)
\;, 
\label{DeltaB}
\eea
in agreement with the formulas in \cite{Hohm:2013uia}. 
Similarly, the covariant gauge variations of the 2-forms read 
 \be
  \Delta_{\Lambda} {\cal B}_{\mu\nu} \ = \ \Lambda\bullet {\cal F}_{\mu\nu}\;, 
 \ee
with the 2-form field strength defined as above. Moreover, recalling the definition (\ref{E7calD}), 
this field strength satisfies the Bianchi identity 
 \be
  D{\cal F}_{(2)} \ = \ {\cal D}{\cal H}_{(3)}\;, 
\label{BianchiE7}
 \ee 
where ${\cal H}_{(3)}\in \frak{g}^*$ is the covariant 3-form field strength of the 2-form gauge field. 
There is a natural topological (Chern--Simons-type) action for the $p$-form gauge fields for $p=1,2,3$. 
It can be written efficiently as a boundary action in five dimensions
in terms of the 2- and 3-form curvatures ${\cal F}_{(2)}\in X_0$ and ${\cal H}_{(3)}\in X_1$, respectively,  
 \be\label{topological}
  S_{\rm top} \ \propto \ \int_{M_5} \, \Theta\big({\cal F}_{(2)}  \stackrel{\wedge}{,}  \, {\cal H}_{(3)}\big)\;, 
 \ee
in terms of the embedding tensor (\ref{E7embedding}).

Having defined the tensor hierarchy up to the level relevant for the present construction we now describe the 
full theory. The bosonic field content is given by 
 \be
  \big\{ g_{\mu\nu}, \, {\cal M}_{MN}, \,A_{\mu}{}^{M},\, {\cal B}_{\mu\nu}\big\}\;, 
 \ee
where $\mu,\nu=0,\ldots, 3$ and all fields depend on external coordinates $x^{\mu}$ and internal 
coordinates $Y^M$. Here $g_{\mu\nu}$ is an E$_{7(7)}$ singlet of density weight 1, 
${\cal M}_{MN}$ is the generalized metric corresponding to the ${\rm E}_{7(7)}/{\rm SU}(8)$ coset space, 
encoding the internal `scalar' degrees of freedom, and $A_{\mu}$, ${\cal B}_{\mu\nu}$ are the gauge 
fields entering the tensor hierarchy. The bosonic action reads 
 \be
  \begin{split}
   S = \int {\rm d}^4x\,{\rm d}^{56}Y\sqrt{g}\;\Big(&\widehat{R}+\tfrac{1}{48}D^{\mu}{\cal M}^{MN} D_{\mu}{\cal M}_{MN} - V(g, {\cal M})\,-\\
   &-\tfrac{1}{8}{\cal M}_{MN}\, {\cal F}^{\mu\nu M} {\cal F}_{\mu\nu}{}^{N} +{\cal L}_{\rm top}\Big)\;, 
  \end{split}
 \ee 
with the Lagrangian ${\cal L}_{\rm top}$ corresponding to the topological action (\ref{topological}). 
The `potential' term is given by 
 \be
  V(g, {\cal M}) \ = \ -{\cal R}-\tfrac{1}{4}{\cal M}^{MN}\nabla_M g^{\mu\nu}\nabla_N g_{\mu\nu}\;, 
 \ee
where we refer to \cite{Godazgar:2014nqa}  for the E$_{7(7)}$ Ricci scalar ${\cal R}$. 
Finally, the above action has to be subjected to a self-duality constraint on the 56 vector fields 
(so the action is really a pseudo-action), 
 \be
  {\cal F}_{(2)}{}^M \ = \ -\frac{1}{2}\,\Omega^{MN}{\cal M}_{NK} \,\star {\cal F}_{(2)}{}^K\;, 
 \ee
where $\star$ denotes Hodge duality in the external, four-dimensional space. 
Note that, thanks to the topological term, the field equations for the 2-forms $B_{\mu\nu M}$ 
are compatible with this constraint (but the duality equations are not fully implied by the
field equations). 

Upon breaking E$_{7(7)}$ to ${\rm GL}(7)$ or ${\rm GL}(6)\times {\rm SL}(2)$, respectively, and solving the section 
constraints accordingly by restricting the fields to only depend on $4+7$ or $4+6$ coordinates, 
the above theory reduces to either $D=11$ or type IIB supergravity in a split formulation
analogous to that of Einstein gravity reviewed in the introduction.

\subsection{E$_{8(8)}$ exceptional field theory}

The construction of a gauge invariant action functional starts from a Chern--Simons theory
that is built from the Leibniz algebra $\frak{g}^*$.
With $\frak{g}^*$-valued vector fields ${\cal A}_\mu$, the covariant non-Abelian field strength reads
\be
  {\cal F}_{\mu\nu} \ \equiv \ \partial_{\mu}{\cal A}_{\nu}-\partial_{\nu}{\cal A}_{\mu} -\big[{\cal A}_{\mu},{\cal A}_{\nu}\big]  
  \ + \  {\cal D}{\cal B}_{\mu\nu}\;,
  \label{FE8}
 \ee
with the bracket  based on (\ref{LeibnizE8Prod}) and the ${\cal D}$ map from (\ref{DBE8}).
It satisfies a Bianchi identity analogous to (\ref{BianchiE8})
\be
  D{\cal F}_{(2)} \ = \ {\cal D}{\cal H}_{(3)}\;, 
\label{BianchiE8}
 \ee 
 with the covariant 3-form field strength ${\cal H}_{(3)}$ whose explicit form
 shall not matter in the following.
A gauge invariant Chern--Simons functional is straightforwardly constructed as the boundary contribution
of a four-dimensional integral
\bea
S_{\rm CS} &\propto& \int_{M_4}  \,\Theta\left({\cal F} \stackrel{\wedge}{,} {\cal F}\right)
\;,
\label{CSE8}
\eea
with the bilinear form from (\ref{varthetaE8abs}) above. Gauge invariance is manifest
while closedness of the integrand follows from (\ref{BianchiE8}) together with (\ref{thD0}).
The same argument shows that the two-forms ${\cal B}_{\mu\nu}$ do actually not explicitly appear
in the action functional (\ref{CSE8}).
Evaluating the bilinear form for a vector field parametrized as
${\cal A}=(A^M, B_M)\in \frak{g}^*$, it takes the explicit form
\bea
S_{\rm CS} &=&
\int {\rm d}^3x\, {\rm d}Y\, \varepsilon^{\mu\nu\rho}\,\Big( 
2\,\partial_{\mu}A_{\nu}{}^{M}\,B_{\rho}{}_M 
-A_{\mu}{}^{N}\partial_NA_{\nu}{}^{M}\,B_{\rho}{}_M\,+
\nonumber\\
&&\qquad~~~~
   +A_{\mu}{}^{M}\partial_N A_{\nu}{}^{N}\,B_{\rho}{}_M
      -f_{KL}{}^N \partial_\mu A_\nu{}^K \partial_N A_\rho{}^L\,-
 \nonumber\\
   &&\qquad~~~~
 -f^M{}_{KP}\,f^{PN}{}_L{} 
   A_{\mu}{}^{K}\partial_N A_{\nu}{}^{L}\,B_{\rho}{}_M\,-
\nonumber\\
&&\qquad~~~~
-\tfrac23\,f^N{}_{KL} \partial_M\partial_N A_{\mu}{}^K A_\nu{}^M A_{\rho}{}^L\,-
\label{CS}
\\
&&
\qquad\qquad
-\tfrac13 \,f_{MKL} f^{KP}{}_Q f^{LR}{}_S\,A_\mu{}^M \partial_P A_\nu{}^Q \partial_R A_\rho{}^S
\Big) 
\;. 
\nonumber
\eea
The full bosonic action of E$_{8(8)}$ ExFT is given by coupling (\ref{CSE8}) to 
an external metric $g_{\mu\nu}$ and scalar fields
parametrizing a matrix ${\cal M}_{MN} \in {\rm E}_{8(8)}/{\rm SO}(16)$ as
 \be
  \begin{split}
   S = \int {\rm d}^4x\,{\rm d}Y\sqrt{g}\;\Big(&\widehat{R}+\tfrac{1}{48}D^{\mu}{\cal M}^{MN} D_{\mu}{\cal M}_{MN}\,+ \\
   & \qquad +\tfrac12\,{\cal L}_{\rm CS} - V(g, {\cal M})\Big)\;, 
  \end{split}
 \ee 
with a gauge invariant `potential' term $V(g, {\cal M})$ constructed in \cite{Hohm:2014fxa}.

Just as for the other ExFT's, upon breaking E$_{8(8)}$ to ${\rm GL}(8)$ or ${\rm GL}(7)\times {\rm SL}(2)$, respectively, and solving the section 
constraints accordingly by restricting the fields to only depend on $4+8$ or $4+7$ coordinates, 
the above theory reproduces either $D=11$ or type IIB supergravity in a split formulation.

\section{Conclusions and open problems}\label{sec:5}

We have reviewed the higher gauge structures of double and exceptional field theory.
Let us finish with a list of open problems: 

\begin{enumerate}[i)]

\item Can one define finite or large generalized diffeomorphisms using an embedding tensor so as to make contact 
with the double field theory results of \cite{Hohm:2012gk} and to find generalizations to exceptional field theory? 
The action of the Lie algebra on which it is based can be integrated directly, so is there a way to similarly 
`integrate' the insertion of $\vartheta$?

\item Related to the above, for generalized Scherk-Schwarz compactifications there is no known systematic  
way to construct the twist matrices, say from the structure constants $X_{MN}{}^{K}$ of the desired gauge algebra. 
Can this problem be solved by using the `infinite-dimensional' embedding tensor reviewed here? 

\item To which extent can these structures, and in particular the invariant action functionals, 
be defined for  infinite-dimensional duality groups E$_{d(d)}$ with $d>8$\,?

\item Can the $\alpha'$-deformed generalized Lie derivatives of double field theory, as in \cite{Hohm:2013jaa}, 
similarly be obtained from an embedding tensor? 
If so, does this give us a hint of how to generalize this to exceptional field theory? 

\item Perhaps most importantly, does this formulation give a hint of how to formulate true, weakly constrained
double and exceptional field theory which would go genuinely beyond the standard supergravities?  

\end{enumerate}

%\bibliographystyle{prop2015}
%\bibliography{refs}

\providecommand{\othercit}{}
\providecommand{\jr}[1]{#1}
\providecommand{\etal}{~et~al.}

\bibliography{allbibtex}

\bibliographystyle{prop2015}

\end{document}